\documentclass{pasa}%

\usepackage{anyfontsize} 
\usepackage{xspace}
\usepackage{natbib}
\usepackage{booktabs}
\usepackage{multirow}
\usepackage{threeparttable}
\usepackage{tikz}     
\usepackage{graphicx} 
\usepackage{subcaption} 
\usepackage{amsmath}  
\usepackage{amssymb}  
\usepackage{xcolor}
\usepackage[normalem]{ulem}
\usepackage{listings} 
\definecolor{codegreen}{rgb}{0,0.6,0}
\definecolor{codegray}{rgb}{0.5,0.5,0.5}
\definecolor{codepurple}{rgb}{0.58,0,0.82}
\definecolor{backcolour}{rgb}{0.95,0.95,0.92}

\lstdefinestyle{mystyle}{
    backgroundcolor=\color{backcolour},   
    commentstyle=\color{codegreen},
    keywordstyle=\color{magenta},
    numberstyle=\tiny\color{codegray},
    stringstyle=\color{codepurple},
    basicstyle=\ttfamily\footnotesize,
    breakatwhitespace=false,         
    breaklines=true,                 
    captionpos=b,                    
    keepspaces=true,                 
    numbers=left,                    
    numbersep=5pt,                  
    showspaces=false,                
    showstringspaces=false,
    showtabs=false,                  
    tabsize=2
}

\newcommand{\hide}[1]{}

\newcommand{\kms}{\ensuremath{\,{\rm km\,s^{-1}}}\xspace}

\newcommand{\cm}{\ensuremath{\,{\rm cm}}\xspace}

\newcommand{\khz}{\ensuremath{\,{\rm kHz}}\xspace}
\newcommand{\mhz}{\ensuremath{\,{\rm MHz}}\xspace}
\newcommand{\ghz}{\ensuremath{\,{\rm GHz}}\xspace}

\newcommand{\degree}{\ensuremath{^\circ}\xspace}

\newcommand{\hi}{{\rm H\,{\footnotesize I}}\xspace}
\newcommand{\hii}{{\rm H\,{\footnotesize II}}\xspace}

\newcommand{\hna}{\ensuremath{{\rm H}{\rm n}\alpha\xspace}}

\newcommand\arcdeg{\mbox{$^\circ$}}%
\newcommand\arcmin{\mbox{$^\prime$}}%
\newcommand\arcsec{\mbox{$^{\prime\prime}$}}%

\newcommand\fs{\mbox{$.\!\!^{\mathrm s}$}}%

\newcommand\fdeg{\mbox{$.\!\!^\circ$}}%
\newcommand\farcm{\mbox{$.\mkern-4mu^\prime$}}%
\newcommand\farcs{\mbox{$.\!\!^{\prime\prime}$}}%

\usepackage{tikz,hyperref}

\definecolor{lime}{HTML}{A6CE39}

\title[Baseline correction for FAST RRLs]{Baseline correction for FAST radio recombination lines: a modified penalized least squares smoothing technique}

\author[B. Liu et al.]{
    Bin Liu$^{1}$\thanks{E-mail: bliu@nao.cas.cn (LB)}, 
    Lixin Wang$^{2}$,
    Junzhi Wang$^{3}$\thanks{E-mail:junzhiwang@gxu.edu.cn (WJZ)},
    Bo Peng$^{1}$\thanks{E-mail: pb@nao.cas.cn (PB)},
    and Hongjun Wang$^{2}$
\affil{$^{1}$CAS Key Laboratory of FAST, National Astronomical Observatories, Chinese Academy of Sciences, 20A Datun Road, Beijing, 100101, People's Republic of China}
\affil{$^{2}$Shaanxi University of Science and Technology, Weiyang University Park, Xi'an 710021, People's Republic of China}
\affil{$^{3}$Guangxi Key Laboratory for Relativistic Astrophysics, School of Physical Science and Technology, Guangxi University,\\ Nanning 530004, People's Republic of China}
}

\jid{PASA}
\doi{10.1017/pas.\the\year.xxx}
\jyear{\the\year}

\usepackage{aas_macros}
\hypersetup{colorlinks,citecolor=blue,linkcolor=blue,urlcolor=blue}

\hypersetup{draft}

\begin{document}

\begin{frontmatter}
\maketitle

\begin{abstract}
A pilot project has been proceeded to map 1 deg$^2$ on the Galactic plane for radio recombination lines (RRLs) using the Five hundred meter Aperture Spherical Telescope (FAST).
The motivation is to verify the techniques and reliabilities for a large-scale Galactic plane RRL survey with FAST aiming to investigate the ionized environment in the Galaxy.
The data shows that the bandpass of the FAST 19 beam L-band is severely affected by radio frequency interferences (RFIs) and standing wave ripples, which can hardly be corrected by traditional low order polynomials.
In this paper, we investigate a series of penalized least square (PLS) based baseline correction methods for radio astronomical spectra that usually contain weak signals with high level of noise.
Three promising penalized least squares based methods, AsLS, arPLS, and asPLS are evaluated.
Adopting their advantages, a modified method named rrlPLS is developed to optimize the baseline fitting to our RRL spectra.
To check their effectiveness, the four methods are tested by simulations and further verified using observed data sets.
It turns out that the rrlPLS method, with optimized parameter $\lambda=2\times10^8$, reveals the most sensitive and reliable emission features in the RRL map.
By injecting artificial line profiles into the real data cube, a further evaluation of profile distortion is conducted for rrlPLS.
Comparing to simulated signals, the processed lines with low signal-to-noise ratio are less affected, of which the uncertainties are mainly caused by the rms noise.
The rrlPLS method will be applied for baseline correction in future data processing pipeline of FAST RRL survey.
Configured with proper parameters, the rrlPLS technique verified in this work may also be used for other spectroscopy projects.
\end{abstract}

\begin{keywords}
radio lines: ISM -- methods: data analysis -- surveys -- ISM: clouds
\end{keywords}
\end{frontmatter}

\section{Introduction} \label{sect:intr}
The current world largest single dish radio telescope, Five hundred meter Aperture Spherical Telescope \citep[FAST,][]{Qiu1998, Nan2011, Qian2020}, was built in late 2016, and started fully operating in early 2020.
Large single dishes have been proven dominantly in observations of radio recombination lines (RRLs) tracing ionized interstellar medium in the Galaxy \citep[][etc.]{Alves2015, Liu2019, Anderson2021}.
With its unprecedentedly high sensitivity, FAST has great potential to study diffuse ionized gases along the Galactic plane using RRLs.
Using the FAST 19 beam L-band receiving system, a pilot observation has been made to image the Galactic plane with RRLs, from which the data will be used to verify the techniques and reliabilities for a large scale Galactic RRL survey.

The modern developments of electronic devices and wireless communication technologies have made the microwave environment more and more lousy for radio telescopes.
Therefore, radio spectroscopy observations are often contaminated by radio frequency interferences (RFIs) and baseline problems, especially in centimeter wavelength.
Although some frequency ranges are protected for astronomical studies, such as the 21\cm neutral hydrogen (\hi) line around 1420\mhz, there are hardly protections for studies which need a wide frequency coverage.
A typical example is the observations of RRLs, whose line rest frequencies cover the entire radio frequency range from $\sim$ 100\ghz to $\sim$ 100\mhz.

Baseline removal is an essential preprocessing step for spectral data analysis. 
The purpose is to remove the artificial baseline structure caused by the electronics or broad RFI features, and to retain the astronomical signal unaffected.
A common way of baseline estimation is to perform a low order ($\leq3$) polynomial least-square fitting.
Since our observations were made during the early operation of FAST, the frequency bandpass were not ideally clean and flat.
Many RRL spectral segments were affected by RFIs and baseline ripples, to which the low order polynomial baseline fitting are mostly in vain.
New method of baseline correction is then indispensable before the line profiles to be accurately fitted.

Based on penalized least squares (PLS) smoothing technique, baseline correction methods have been developed and applied in Raman and infrared spectroscopic analysis.
The basic idea of PLS is to balance between fidelity to the original data and the roughness of the fitted baseline by combining least squares smoothing together with a penalty on roughness of an estimation.
The PLS algorithm for baseline correction was first introduced by \citet{Eilers2003, Eilers2004} and named as asymmetric least squares (AsLS).
To improve the results of baseline correction, several modified methods inspired by AsLS have been developed subsequently, they are: 1) adaptive iteratively reweighted penalized least squares \cite[airPLS,][]{Zhang2010}; 2) improved asymmetric least squares \cite[IAsLS,][]{He2014}; 3) asymmetrically reweighted penalized least squares \cite[arPLS,][]{Baek2015}; and 4) adaptive smoothness parameter penalized least squares \cite[asPLS,][]{Zhang2020}.

In this paper, we focus on the application and evaluation of the PLS-based baseline fitting algorithms applied to the RRL spectra obtained with FAST.
Section\,\ref{sect:obs} describes the RRL observations made by FAST and the data reduction pipeline for the spectral line imaging.
Section \,\ref{sect:pls} reviews the theory of the existing PLS-based algorithms and introduces our modified method for FAST RRL data, rrlPLS.
Section\,\ref{sect:simu} presents the simulation work, where the AsLS, arPLS, asPLS, and rrlPLS are evaluated using simulated data set and the optimized parameters are listed.
Their applications to actual observed data are shown in Section\,\ref{sect:rrl}.
In Section\,\ref{sect:fake}, we verify the rrlPLS method using the real data cube with artificial line profiles injected.
Conclusions are given in Section\,\ref{sect:con}.

\section{RRL observation and data reduction using FAST}\label{sect:obs}
\subsection{The RRL observation}

The pilot project covers a field of 1 deg$^2$ along the Galactic plane, which was observed using the FAST Multi-Beam On-The-Fly (MBOTF) mode.
This field centering at $l=34\fdeg5, b=0\fdeg0$ is chosen for it contains active star-formation regions, thus intensive RRLs both from discrete \hii regions and diffuse ionized gas are expected.
A reference position off the Galactic plane is adopted for bandpass calibration.
Table\,\ref{tab:obs_conf} gives the detailed information of the targeted region.

MBOTF observations are deployed in the Equatorial system.
We scan the targeted region twice in each session, along RA and Dec axis respectively, with a scan speed of $\sim$33\arcsec per second.
The offset position was observed for 5 minutes before and after each MBOTF mapping and a flux calibrator was observed at the beginning to confirm the power stability of the noise diode during different sessions.
Flux calibration was done adopting the temperature of the noise diode provided by the FAST official website.
To summarize, the observing procedure for each session is: 1) flux calibrator; 2) offset position; 3) MBOTF in RA; 4) offset position; 5) MBOTF in Dec; 6) offset position. 

\begin{table}
    \caption{ The sky coverage for the 1 deg$^2$ RRL mapping.}
    \label{tab:obs_conf}
    \centering
    \begin{tabular}{ll}
        \hline\hline
    Observing Area & Values \\
    \hline
    Map center (Gal) & $l=34\fdeg5, b=0\fdeg0$ \\
    Map center (Equ)    & 18$^{\rm{h}}$54$^{\rm{m}}$17$\fs$95+01$\degree$23$\arcmin$43$\farcs$9 \\
    RA range (J2000)          & 18$^{\rm{h}}$52$^{\rm{m}}$42$\fs$58$\sim$18$^{\rm{h}}$56$^{\rm{m}}$16$\fs$26 \\
    Dec range                 & +00$\arcdeg$35$\arcmin$24$\farcs$5$\sim$+01$\arcdeg$56$\arcmin$10$\farcs$0\\
    OFF position              & 18$^{\rm{h}}$48$^{\rm{m}}$02$\fs$57+01$\arcdeg$48$\arcmin$44$\farcs$2\\
    \hline
\end{tabular}
\end{table}

The frequency bandpass of the FAST L-band is from 1050 to 1450\mhz, which covers twenty hydrogen $\alpha-$RRLs from H165$\alpha$ to H184$\alpha$.
The spectrometer records one spectrum per second which covers a digital bandwidth of 500\mhz with with 2$^{20}$ channels resulting the frequency resolution of $\sim$0.478\khz.
The corresponding velocity resolutions of the twenty RRL segments are from 0.099 to 0.137\,\kms.

\begin{table}
    \caption{ The backend configuration. }
    \label{tab:backend}
    \centering
    \begin{tabular}{ll}
        \hline\hline
    Parameters & Values \\
    \hline
    Targeted RRLs (\hna)    & H165$\alpha$$-$H184$\alpha$ \\
    Frequency Range             & 1050$-$1450\mhz            \\
    Digital Bandwidth           & 500\mhz                    \\
    Number of Channels          & 2$^{20}$ (1\,M)            \\
    Frequency Resolution        & 0.478\khz                  \\
    Velocity Resolution Range   & 0.099$-$0.137\,\kms        \\
    Integration per Sample      & 1\,s                       \\
    \hline
\end{tabular}
\end{table}

\subsection{Data reduction}

A data reduction pipeline has been developed to process the FAST spectra from MBOTF observations.
Three major steps are applied including radio frequency interference (RFI) excision, calibration, and baseline removal.
After calibration, the full bandpass are cut into individual RRL segments, to which the baseline removal is deployed.
The system properties adopted for calibration are given in \citet{jiang2020}.

The frequency channels affected by strong and broad RFIs, which may come from satellites, ground radar or communication stations, are firstly flagged out. 
To excise weak, narrow, and transitory RFIs, a median absolute deviation filter is applied \citep{Liu2019}, with a window width of 25 channels and intensity threshold above 3 times of the spectral rms.

\begin{figure}
\centering
\includegraphics[width=0.5\textwidth, angle=0]{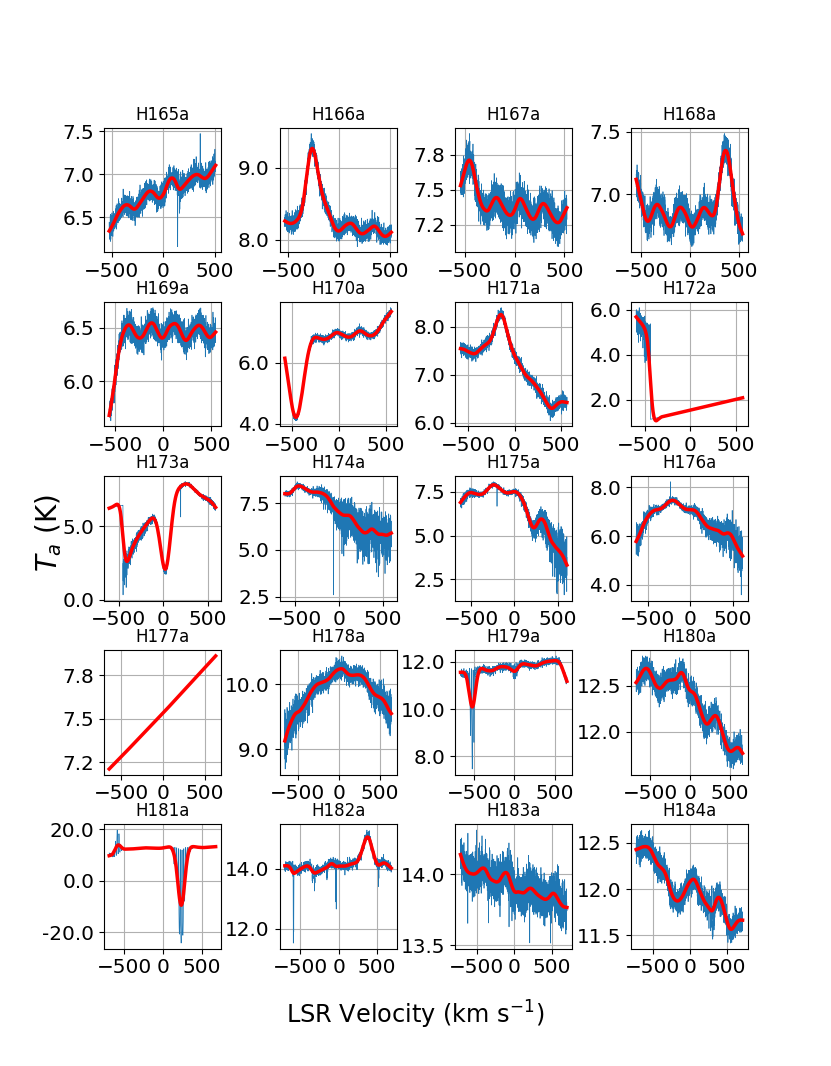}
\caption{The averaged spectra of RRL segments over a 60\,seconds OTF scan.
    The blue lines are the spectra and the red lines are the results of the asymmetric least squares smoothing (AsLS).
    In the pipeline, baseline removal was applied to the raw spectrum with 1\,second dumping time. 
    We show the averaged spectra only for the purpose of illustration since the baseline features are hard to be seen from the individual spectrum.}
\label{fig:rrlseg}
\end{figure}

The bandpass of the FAST 19 beam L-band receiver is affected by standing wave ripples with a typical width of $\sim$100\,\kms \citep[see][]{jiang2020}.
Figure\,\ref{fig:rrlseg} shows the averaged baseline of the twenty segments over 60 seconds.
We show the averaged spectra in Figure\,\ref{fig:rrlseg} only for a better illustration of the baseline features.
In the pipeline, the baseline removal was applied to the raw spectrum with 1\,second integral time. 
Automatic polynomial or sinusoid fitting could not deal with such unstable baseline situations.

As a test, AsLS was applied in the pipeline, which was the first PLS based methods originally developed for baseline correction in Chemistry and Raman spectroscopy \citep{Eilers2004, Peng2010, Zhang2020b}.
Differing from its original application, where both the baseline ripple and the spectral line intensity are strong while the noise are negligible, in our data the baseline ripples and the noise are significant but the spectral line signals are usually weak.
The red lines in Figure\,\ref{fig:rrlseg} illustrate the result of the AsLS test.
For our pipeline, a optimized PLS-based method, rrlPLS, was finally adopted, which is introduced in Section\,\ref{sect:pls:rrlpls}.

The spectra of individual RRL are spatially re-sampled and grided into data cube with 1$'$ pixel size ($\sim1/3$ beam size), with a Gaussian kernel following the instruction given by \citet{Mangum2007}.
In each observing session, one data cube is created for each RRL segment from the combined data sets of the two MBOTF scans. 
The cubes for the same RRL segment from different sessions are then averaged.
Finally, we stack the data cubes of all segments in order to achieve a high signal-to-noise ratio.
Since the beam size of a telescope varies with frequency, the spatial resolutions are different over those twenty RRLs.
Before stacking, the cubes of different lines are convolved to an uniform beam size of \textcolor{red}{\sout{3\farcm3} 3\farcm4}, which is the FAST \textcolor{red}{Half Power Beam Width (HPBW) \sout{beam size} at \sout{1.1\ghz} 1050\mhz} (near the rest frequency of \textcolor{red}{\sout{H181$\alpha$} H184$\alpha$)}.

\section{The PLS-based methods for baseline correction}\label{sect:pls}
As the first PLS-based baseline fitting method, the AsLS was proposed by \citet{Eilers2003, Eilers2004} and has proved effective.
Since then several improved PLS-based algorithms have been developed including airPLS \citep{Zhang2010}, IAsLS \citep{He2014}, arPLS \citep{Baek2015}, and asPLS \citep{Zhang2020}.
arPLS and asPLS were designed to deal with noisy spectrum, they are described and discussed bellow in details along with the AsLS and our modified method rrlPLS.
No further analysis is applied to the airPLS and IAsLS methods since our test with these two did not present effective baseline fitting results to spectra with high noise level.

\subsection{The AsLS method}\label{sect:pls:asls}

To consider a power spectrum with length of $m$ obtained by a radio telescope, its vector model $\mathbf{y} = [y_1, y_2, \cdots, y_i, \cdots, y_m]^{\mathrm{T}}$ is a composition of the profile of spectral line $\mathbf{s} = [s_1, s_2, \cdots, s_i, \cdots, s_m]^{\mathrm{T}}$, a baseline vector $\mathbf{b} = [b_1, b_2, \cdots, b_i, \cdots, b_m]^{\mathrm{T}}$, and random noise $\mathbf{n} = [n_1, n_2, \cdots, n_i, \cdots, n_m]^{\mathrm{T}}$, which gives
\begin{equation}\label{equ:y}
\mathbf{y=s+b+n}.
\end{equation}

Based on the Whittaker smoother \citep{Eilers2003}, \cite{Eilers2004} proposed the function to be minimized for a smoothing background,
\begin{equation}\label{equ:q}
    Q = \sum_{i=1}^{m} w_i \left(y_i-b_i\right)^2 +\lambda \sum_{i=1}^{m}\left(\Delta^2 b_i\right)^2.
\end{equation}
$\Delta$ is the first-order difference and $\Delta^2$ stands for the second-order difference, which gives
\begin{equation}\label{equ:d}
    \begin{aligned}
    \Delta^2 b_i = \Delta(\Delta b_i) 
             = (b_i - b_{i-1}) - (b_{i-1} - b_{i-2}) \\
             = b_i - 2b_{i-1} + b_{i-2}.
   \end{aligned}
\end{equation}
The weight vector $\mathbf{w} = [w_1, w_2, \cdots, w_i, \cdots, w_m]^{\mathrm{T}}$ are chosen asymmetrically according to
\begin{equation} \label{equ:w}
    w_i = 
    \begin{cases}
        p,   & y_i > b_i\\
        1-p, & y_i \le b_i
    \end{cases},
    (0<p<1).
\end{equation}
$p$ and $\lambda$ are smoothing parameters which should be optimized based on the data properties and preset by the user.

For convenience of implementation in programming and to simplify the equations, we adopt the form of linear algebra.
Let $\mathbf{W}$ to be $m\times m$ diagonal matrix with $\mathbf{w}$ on its diagonal
\begin{equation}\label{equ:W}
    \mathbf{W} = 
    \begin{bmatrix}
        w_1    &  0     & \cdots & 0       \\
        0      &  w_2   & \cdots & 0       \\
        \vdots & \vdots & \ddots & \vdots  \\
        0      & 0      & \cdots & w_m     \\
    \end{bmatrix},
\end{equation}
and $\mathbf{D}$ as the $(m-2)\times m$ matrix such that $\mathbf{Db}=\Delta^2\mathbf{b}$.
According to Equation\,\ref{equ:d},
\begin{equation}\label{equ:D}
    \mathbf{D} = 
    \begin{bmatrix}
        1      & -2     & 1      & 0      & \cdots & 0      & 0      & 0      \\
        0      &  1     & -2     & 1      & \cdots & 0      & 0      & 0      \\
        \vdots & \vdots & \vdots & \vdots & \ddots & \vdots & \vdots & \vdots \\
        0      & 0      & 0      & 0      & \cdots & 1      & -2     & 1      \\
    \end{bmatrix}.
\end{equation}
Thus Equation\,\ref{equ:q} can be rewritten to
\begin{equation}
    Q = \mathbf{(y-b)}^{\mathrm{T}}\mathbf{W(y-b)} + \lambda\mathbf{b}^{\mathrm{T}}\mathbf{D}^{\mathrm{T}}\mathbf{Db},
\end{equation}

By finding the vector of partial derivatives and equating it to zero
\begin{equation}\label{equ:dq}
    \frac{\partial Q}{\partial \mathbf{b}^{\mathrm{T}}} = -2\mathbf{W}(\mathbf{y-b}) + 2\lambda \mathbf{D}^{\mathrm{T}}\mathbf{Db} = 0,
\end{equation}
\begin{equation} \label{equ:b}
    (\mathbf{W}+\lambda\mathbf{D}^{\mathrm{T}}\mathbf{D})\mathbf{b} = \mathbf{Wy}.
\end{equation}
Solving Equation\,\ref{equ:b}, we will obtain the optimal solution of baseline $\mathbf{b}$.

Difficulty lies in choosing values of $p$ and $\lambda$ objectively for the AsLS method.
Experience has shown that this algorithm, using visual inspection to choose the parameters $p$ and $\lambda$ is effective and fast.
For a baseline estimate, $p$ near zero and rather large $\lambda$ make $\mathbf{b}$ follow the valleys of $\mathbf{y}$, i.e. $p=0.001$ and $\lambda = 10^5$.

To start the calculation, the initial weights have to be assigned.
Thus $w_i=1$ is set to obtain an initial baseline $\mathbf{b_0}$, which is then adopted to derive new weights.
Multiple literation are then followed to update the weight vector $\mathbf{w}$ and to estimate better baseline $\mathbf{b}$.
The converging solution will be reached quickly and reliably in about 10 iterations.

\subsection{arPLS}\label{sect:pls:arpls}

In order to perform baseline correction in noisy environment, \citet{Baek2015} proposed the arPLS algorithm.
Given the optimizing equation Equation\,\ref{equ:b} from AsLS, they assign the weight vector $\mathbf{w}$ according to the following equation:
\begin{equation} \label{equ:w2}
    w_i = 
    \begin{cases}
        \mathrm{logistic}\left(y_i-b_i, m_{\mathrm{d}^-},\sigma_{\mathrm{d}^-}\right),   & y_i > b_i\\
        1, & y_i \le b_i
    \end{cases}
\end{equation}
where $m_{\mathbf{d^-}}$ and $\sigma_{\mathbf{d^-}}$ are the mean and standard deviation of $\mathbf{d^-}$.
Defined as $\mathbf{d = y - b}$, $\mathbf{d^-}$ is the negative values of $\mathbf{d}$ when $y_i<b_i$.
The logistic function is introduced as follows:
\begin{equation} \label{equ:log}
    \mathrm{logistic}(d,m,\sigma) = \frac{1}{1+e^{k(d-(-m+s\sigma))/\sigma}},
\end{equation}
where $k$ and $s$ are asymmetric and shifting coefficients, which can be used to squeeze the transient region and to shift the weight curve along x-axis.
The default values given by \citet{Baek2015} is $k=2$ and $s=2$.

\subsection{asPLS}\label{sect:pls:aspls}

In order to attenuate the baseline boost at line peaks, \citet{Zhang2020} proposed the asPLS method.
With the increase of $\lambda$, the smoothed curve in the line peak region is closer to the actual baseline, while in line free regions the curve deviates further from baseline.
Their idea is to adopt different smoothing parameter $\lambda$ for different channels of the spectrum, meaning that to set large $\lambda$ in line peak regions and small value in line free regions.

To implement the asPLS algorithm, a coefficient vector {\boldmath$\alpha$} is introduced to tune the amplitude of $\lambda$.
The minimizing equation, Equation\,\ref{equ:q}, can then be re-written as
\begin{equation}\label{equ:q2}
Q = \sum_{i=1}^{m} w_i \left(y_i-b_i\right)^2 + \sum_{i=1}^{m}(\alpha_i \lambda)\left(\Delta^2 b_i\right)^2,
\end{equation}
where $\alpha_i$ follows
\begin{equation}\label{equ:alpha}
    \alpha_i = \frac{\mathrm{abs} \left(y_i - b_i \right)}{\mathrm{max}\left(\mathrm{abs}\left(\mathbf{y-b}\right)\right)},
\end{equation}
where $\mathrm{abs}()$ is to calculate the absolute value and $\mathrm{max}()$ is to find the maximum value.
According to Equation\,\ref{equ:alpha}, a large value of $\alpha_i$ is given in the line peak region where the difference between $\mathbf{y}$ and $\mathbf{b}$ is large.
And small $\alpha_i$ are introduced in line free regions.
\citet{Zhang2020} introduced the weight function for asPLS following
\begin{equation} \label{equ:w3}
    w_i = \frac{1}{1+ e^{k \left( d_i-\sigma_{\mathrm{d}^-}\right)/\sigma_{\mathrm{d}^-}}},
\end{equation}
where $k$ is asymmetric coefficient with a default value of 2.

\subsection{A modified method: rrlPLS}\label{sect:pls:rrlpls}

The arPLS and asPLS methods both introduced pros and cons for baseline estimations compared to AsLS.
In order to optimize the fitting results to the real RRL data observed with FAST, a modified method is introduced by combining the features of arPLS and asPLS, which is named as modified penalized least square for FAST radio recombination lines (rrlPLS).

As is described in Section\,\ref{sect:obs}, the observation of RRL mapping with FAST uses MBOTF mode.
The raw spectra, which are recorded with a changing pointing, have to be processed directly.
Averaging is not an option until data cubes are being produced during re-gridding.
Thus in the raw data to be processed, RRL signals are commonly weak and accompanied by relatively high noise.

\begin{figure}
\centering
\includegraphics[width=0.4\textwidth, angle=0]{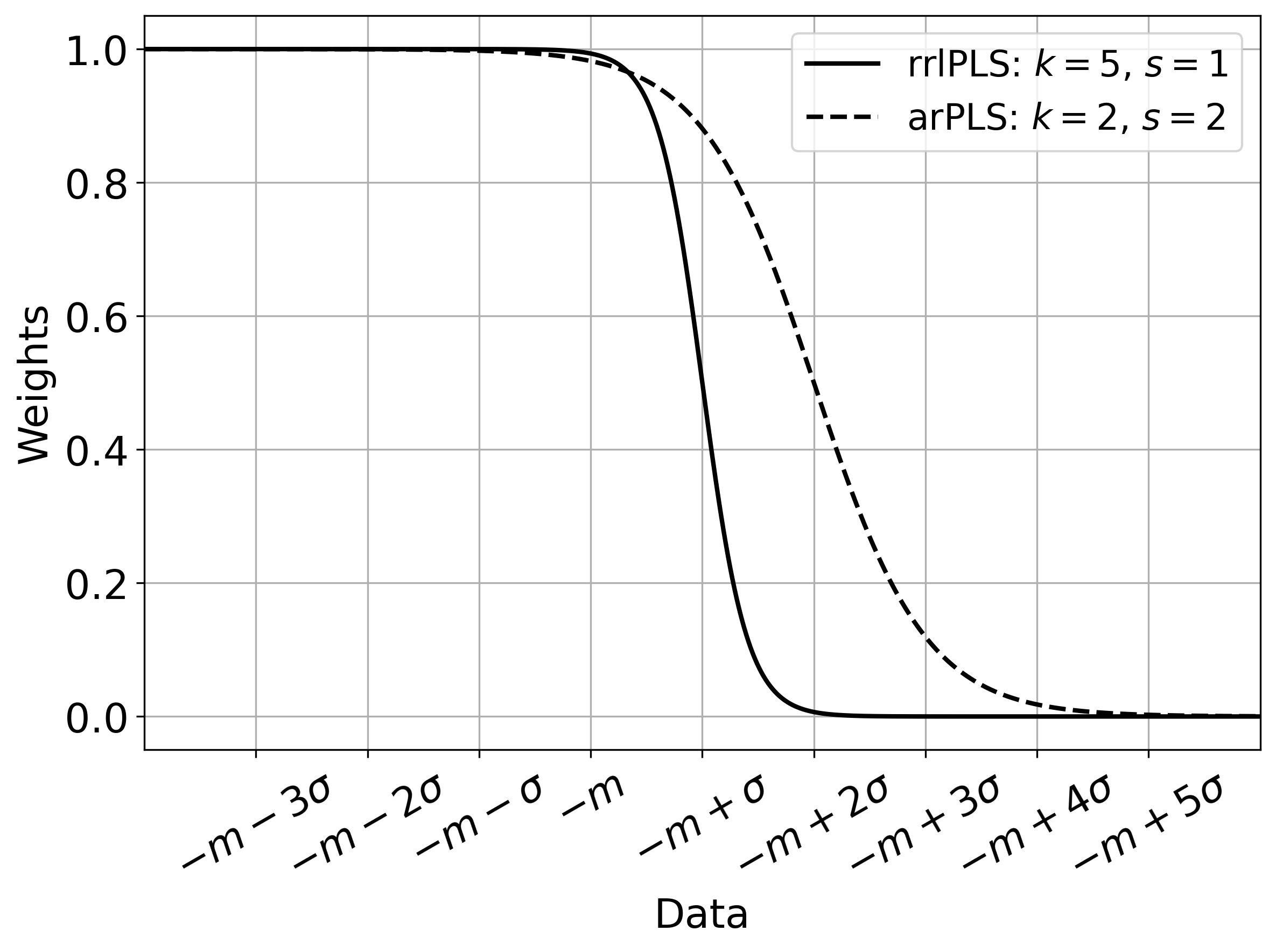}
\caption{
    The weight curve for rrlPLS (solid line) with k=5, s=1 and the default weight curve of arPLS (dashed line) with k=2, s=2.
}\label{fig:rrlpls_w}
\end{figure}

In our modified method, a re-shaped weight function is derived from Equation\,\ref{equ:log}, where the asymmetric coefficient is set to $k=5$ and shifting coefficient \sout{as} $s=1$.
Comparing to the default values of arPLS, the new curve assigns smaller weights to positive differences and follows a more sharp trend on the negative side (see Figure\,\ref{fig:rrlpls_w}).

Meanwhile, we adopt the idea of setting different $\lambda$ with the {\boldmath$\alpha$} according to Equation\,\ref{equ:alpha}.
Flatter baseline is obtained to the line peak region with larger $\lambda$, whereas smaller $\lambda$ produces more curvy baseline for the line free regions.
Since the weight curve is fixed, it remains only one parameter, $\lambda$, to be optimized. 

\section{Investigations with simulated spectra}\label{sect:simu}

The function of PLS-based methods for baseline correction strongly depends on line intensities, noise level, and the amplitude of baseline ripple. 
Also the position of the line on the baseline ripple (or the `phase' of the standing wave where the line is located) affects the fitting results significantly.
For an unbiased comparison between all the methods, we perform a simulation by varying all the related conditions to obtain a statistically significant conclusion.

\subsection{The simulation configuration}\label{sect:simu:config}

The spectra for simulation are generated following Equation\,\ref{equ:y}.
To match with the RRL spectra given by the FAST pipeline, the local standard of rest (LSR) velocity range is from $-$400 to $+$400\,\kms with a resolution of 0.5\,\kms. 
Accordingly, the length of the spectral vector is 1600.
The line profile $\mathbf{s}$ is modeled with a Gaussian function, whose amplitude is 1 as the relative line peak intensity and FWHM is 20\,\kms for the typical line width of Galactic RRLs.
To imitate standing wave ripples in the frequency bandpass of FAST, the baseline vector $\mathbf{b}$ is modeled by a sinusoidal function with a period of 200\,\kms in velocity.

Considering that the RRL intensities vary from sources, different baseline and noise conditions are configured.
We define the signal-to-baseline ratio R$_{\mathrm{b}}$ as the ratio of the line peak intensity (always equals to 1) to the amplitude of sine wave of baseline.
Random noise with Normal distribution is added to the spectral model according to the pre-set signal-to-noise ratio R$_{\mathrm{n}}$, which is the ratio of the line peak intensity to the standard deviation of the noise vector $\mathbf{n}$.
Spectra are then simulated with two different pairs of R$_{\mathrm{b}}$ and R$_{\mathrm{n}}$ for different case studies:
\begin{description}
    \item[\bf{Case A} ($\mathrm{R_{b}}=5, \mathrm{R_{n}}=5$)] This is an ideal case consisting a clear detection of strong line with weak baseline ripples. 
    \item[\bf{Case B} ($\mathrm{R_{b}}=3, \mathrm{R_{n}}=3$)] This is a difficult scenario, in which the line signal is relatively weak due to intensive noise level and strong baseline ripples.
\end{description}

Baseline fittings with the four methods introduced in Section\,\ref{sect:pls} to the simulated spectra are then performed.
For each case, multiple spectra are simulated with 200 different line peak velocities from $-$100 to $+$100\,\kms and with random noise generated from 50 different seeds for each velocity.
Thus, 10,000 tests are conducted for each pair of parameters for each method.

Two factors are introduced to examine the fitting results.
One is the relative loss of the line peak intensity, which is defined as
\begin{equation}\label{equ:loss}
    loss = \frac{F_{\mathrm{fit}} - F_{\mathrm{sim}}}{F_{\mathrm{sim}}} \times 100\%,
\end{equation}
where $F_{\mathrm{fit}}$ is the fitted line peak intensity of the corrected spectrum and $F_{\mathrm{sim}}$ is the original line peak intensity for simulation.
The astronomical spectra are always noisy and the spectral line intensities are normally weak.
The fitted baseline is usually overestimated in line peak regions when the noise level is high, thus the flux loss of line peak intensity is introduced.
Similarly we also examine the relative deterioration of the spectral rms noise, which is defined as
\begin{equation}\label{equ:dete}
    deterioration = \frac{\sigma_{\mathrm{res}} - \sigma_{\mathrm{noi}}}{\sigma_{\mathrm{noi}}} \times 100\%,
\end{equation}
where $\sigma_{\mathrm{res}}$ is rms of residual of the corrected spectrum after removing the fitted line profile, and $\sigma_{\mathrm{noi}}$ is rms of the simulated noise.
Better baseline removal causes smaller rms deterioration.

For stable and reliable baseline fitting, the standard deviation of the distribution of the two factors should be small and the mean should be close to zero.
The distribution of those two factors are evaluators in the procedures of parameter optimization.
In order to obtain the optimized values of each method, we manually iterate over the parameter space with small steps to approach the values that yield the best results.

\subsection{Results}\label{sect:simu:result}
Table\,\ref{tab:results} summarizes the fitting results with optimized values of parameters for conditions of both Case A and B.
The details of results of the four PLS-based methods are discussed as follows.
\begin{table*}
    \caption{The summary table of opimized smulation results for AsLS, arPLS, asPLS, and rrlPLS methods.}
\label{tab:results}
\centering
\begin{threeparttable}
    \begin{tabular}{ccccccccccc}
        \hline\hline
        \multicolumn{3}{c}{Simulation conditions}                             & Methods                & \multicolumn{4}{c}{Optimized Parameters}                                                       & \multicolumn{3}{c}{Fitting results} \\
        \hline
        Case               & R$_{\rm{b}}$       & R$_{\rm{n}}$       &                       & $\lambda$                       & $p$                   & $k$                 & $s$                  & Factor & $\mu (\%)$ & $\sigma (\%)$ \\
        \hline
        \multirow{8}{*}{A} & \multirow{8}{*}{5} & \multirow{8}{*}{5} & \multirow{2}{*}{AsLS} & \multirow{2}{*}{$5\times10^{5}$}&\multirow{2}{*}{0.03}& \multirow{2}{*}{-} & \multirow{2}{*}{-} & Loss & $-$9.6 & 3.3 \\
                           &                    &                    &                       &                                 &                     &                    &                    & Deter.& 1.5 & 0.6 \\
                           &                    &                    & \multirow{2}{*}{arPLS} & \multirow{2}{*}{$1\times10^{6}$}&\multirow{2}{*}{-}& \multirow{2}{*}{-} & \multirow{2}{*}{-} & Loss & $-$8.0 & 2.7 \\
                           &                    &                    &                        &                                 &                     &                         &               & Deter.& $-$0.3 & 0.2 \\
                           &                    &                    & \multirow{2}{*}{asPLS} & \multirow{2}{*}{$5\times10^{5}$}&\multirow{2}{*}{-}& \multirow{2}{*}{-} & \multirow{2}{*}{-} & Loss & $-$2.2 & 6.8 \\
                           &                    &                    &                        &                                 &                     &                         &               & Deter.& 7.8 & 5.2 \\
                           &                    &                    & \multirow{2}{*}{rrlPLS} & \multirow{2}{*}{$1\times10^{7}$}&\multirow{2}{*}{-}& \multirow{2}{*}{5} & \multirow{2}{*}{1} & Loss & $-$3.3 & 3.8 \\
                           &                    &                    &                        &                                 &                     &                         &               & Deter.& $-$0.2 & 0.3 \\
        \hline
        \multirow{8}{*}{B} & \multirow{8}{*}{3} & \multirow{8}{*}{3} & \multirow{2}{*}{AsLS} & \multirow{2}{*}{$5\times10^{5}$}&\multirow{2}{*}{0.03}& \multirow{2}{*}{-} & \multirow{2}{*}{-} & Loss & $-$11.5 & 5.6 \\
                           &                    &                    &                       &                                 &                     &                         &               & Deter.& 1.3 & 0.5 \\
                           &                    &                    & \multirow{2}{*}{arPLS} & \multirow{2}{*}{$1\times10^{6}$}&\multirow{2}{*}{-}& \multirow{2}{*}{-} & \multirow{2}{*}{-} & Loss & $-$16.8 & 4.4 \\
                           &                    &                    &                        &                                 &                     &                         &               & Deter.& $-$0.2 & 0.2\\
                           &                    &                    & \multirow{2}{*}{asPLS} & \multirow{2}{*}{$5\times10^{5}$}&\multirow{2}{*}{-}& \multirow{2}{*}{-} & \multirow{2}{*}{-} & Loss & $-$4.7 & 10.6 \\
                           &                    &                    &                        &                                 &                     &                         &               & Deter.& 7.5 & 4.9\\
                           &                    &                    & \multirow{2}{*}{rrlPLS} & \multirow{2}{*}{$1\times10^{7}$}&\multirow{2}{*}{-}& \multirow{2}{*}{5} & \multirow{2}{*}{1} & Loss & $-$6.6 & 6.3\\
                           &                    &                    &                        &                                 &                     &                         &               & Deter.& $-$0.1 & 0.3 \\
        \hline
    \end{tabular}
    \begin{tablenotes}
    \item Col. 1-3 show the conditions of simulated spectra following the description given in Section\,\ref{sect:simu}.
     Col. 4 lists the names of PLS-based methods.
     Col. 5-8 are the optimized values of parameters for each method. `-' is marked if not applied.
     Col. 9-11 give the simulation results for the two factors defined by Equation\,\ref{equ:loss} and \ref{equ:dete}. $\mu$ and $\sigma$ are the mean and standard deviation of the results of 10,000 tests for one method under each condition.
    \end{tablenotes}
\end{threeparttable}
\end{table*}

\subsubsection{AsLS}\label{sect:simu:resu:asls}

The smoothing parameter $\lambda$ and weighting parameter $p$ of AsLS are configured within $10^2 < \lambda < 10^9$ and $0.001 < p< 0.5$ to suit for different conditions as suggested.
When $p=0.5$, the algorithm is actually the Hodrick-Prescott filtering algorithm \citep{Hodrick1997} that is widely used for macroeconomic time series.

\begin{figure}
\centering
\includegraphics[width=0.42\textwidth, angle=0]{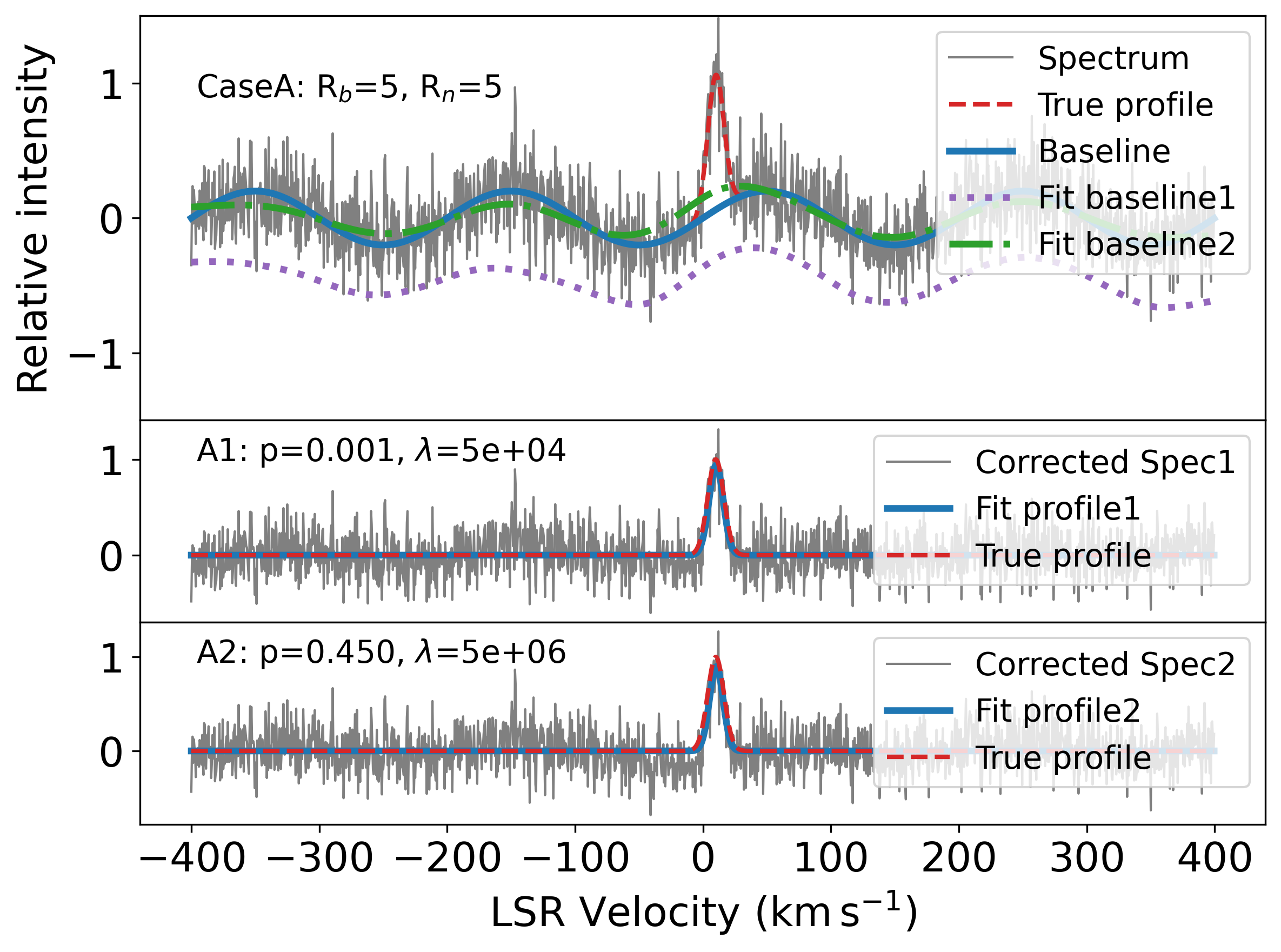}
\caption{
The simulated spectrum and AsLS fitting results under Case A condition.
The top panel shows the simulated spectrum (solid grey), which is the combination of a Gaussian peak (dashed red) as the line profile, a sine wave (solid blue) as the baseline ripple, and white noise.
AsLS baseline fitting results from two different parameter configurations are also plotted (dotted blue and dash-dotted green).
The middle and bottom panel give the baseline corrected spectra (solid grey) from two different parameter configurations which are overlaid by their fitted Gaussian line profiles (solid blue).
The simulated Gaussian peaks (dashed red) are also shown for comparison.
}\label{fig:caseA}
\end{figure}

\begin{figure}
\centering
\includegraphics[width=0.42\textwidth, angle=0]{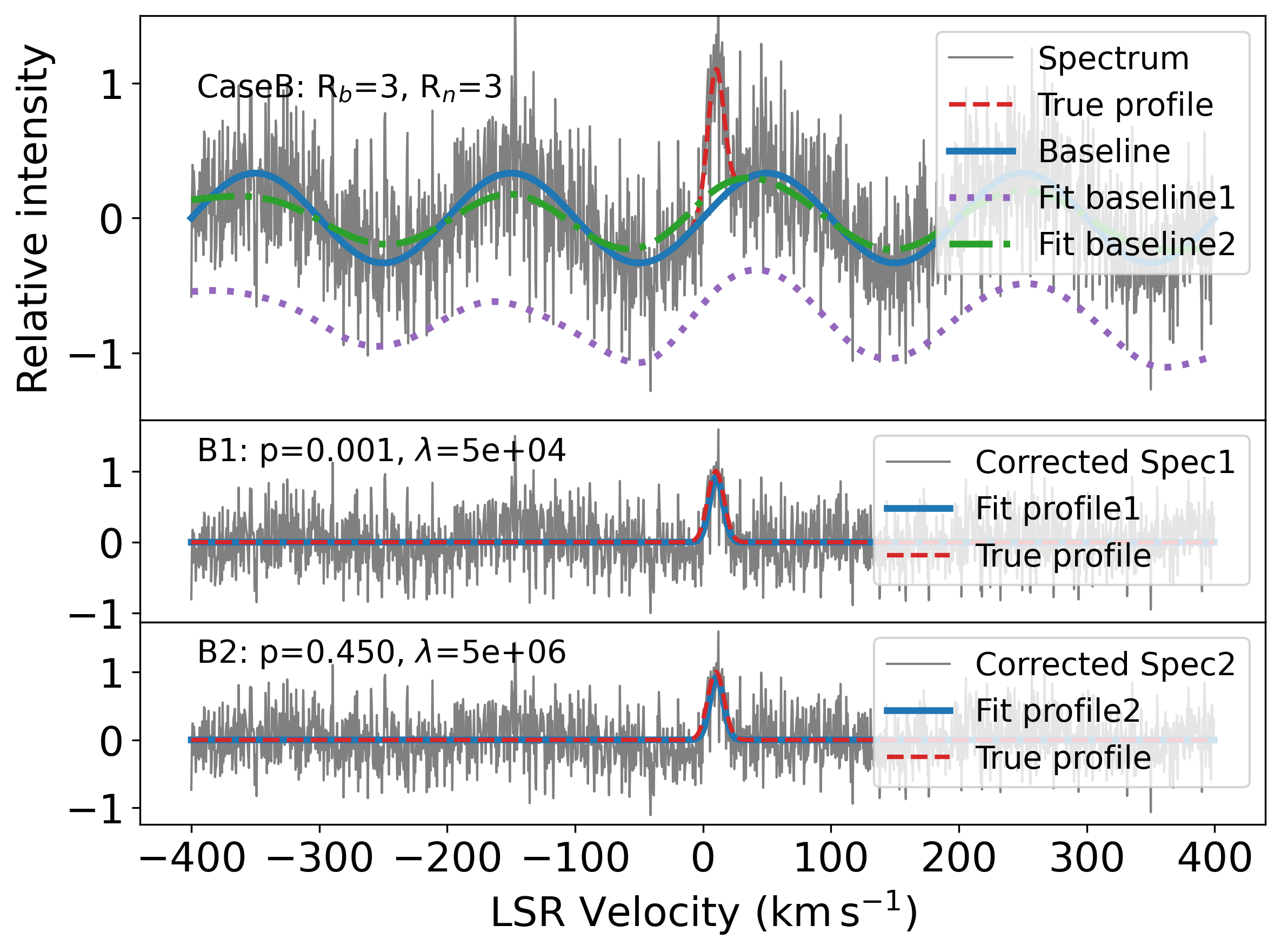}
\caption{
The simulated spectrum and AsLS fitting results under Case B condition.
The plots of the three panels are following the same instruction given in Figure\,\ref{fig:caseA}.
}\label{fig:caseB}
\end{figure}

In order to demonstrate the utility of AsLS, we present two fitting examples with two sets of parameters for both Case A and B.
The two pairs of $\lambda$ and $p$ are $\lambda=5\times 10^4$, $p=0.001$ and $\lambda=5\times10^6$, $p=0.45$. 
The fitting results are presented in Figures\,\ref{fig:caseA} and \ref{fig:caseB}, both of which contain three panels.
The top panel shows the original spectrum (gray) with the simulated line profile (red) and baseline (blue) overlaid.
For comparison, the two fitted baselines are also plotted in this panel.
The middle and bottom plots are the corrected spectra of the two fittings with the simulated and fitted line profiles.
Affected by the noise level, the fitted baseline is likely to be apart from the `real' baseline with a negative offset, especially when $p<<0.5$.
So one more step to correct the spectrum, after removing the fitted baseline, is to further remove the median value of the subtraction.


\begin{figure}
\centering
\includegraphics[width=0.4\textwidth, angle=0]{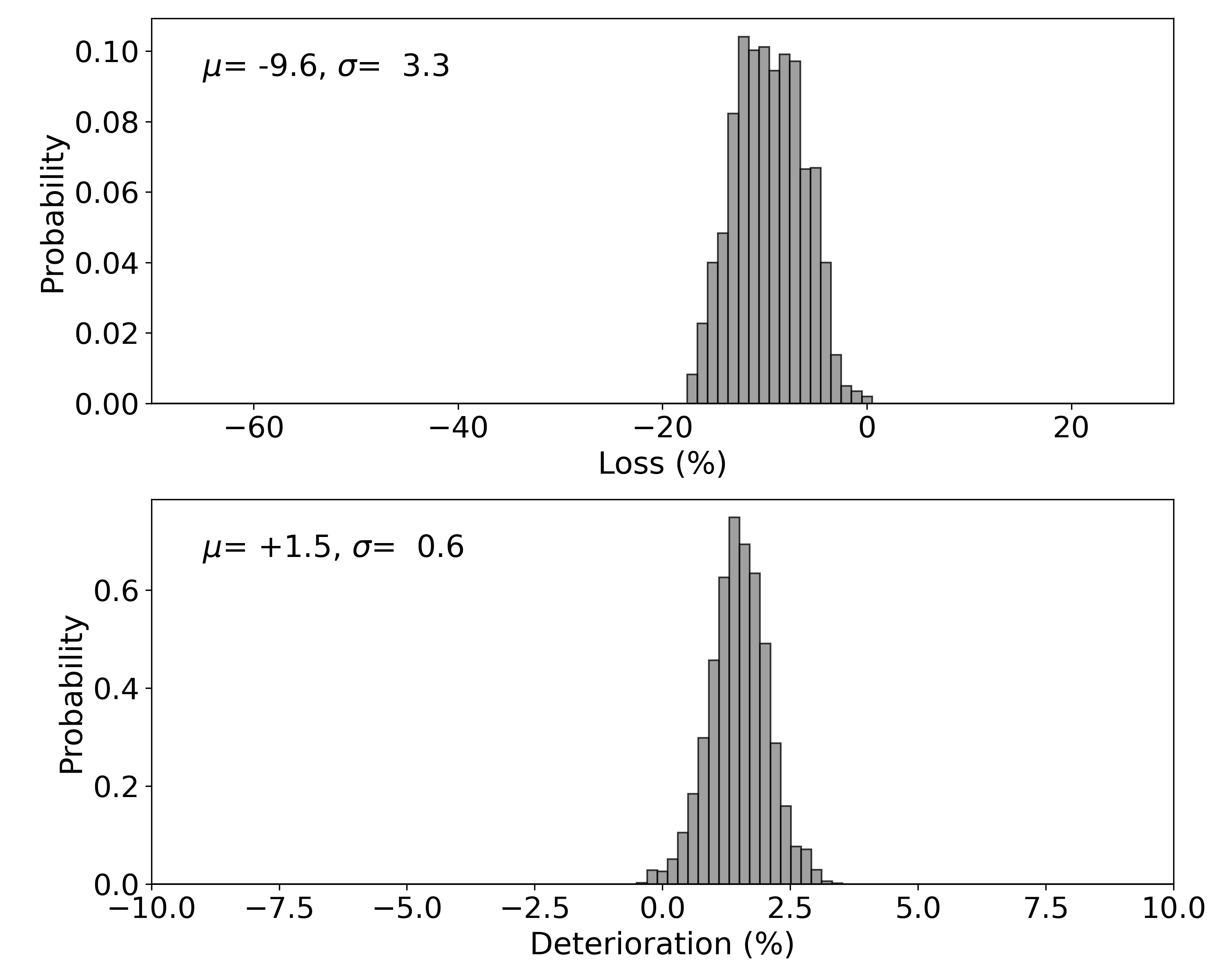}
\caption{
    The distribution of simulation results for Case A using AsLS method with optimized parameters.
    The optimized parameters of AsLS method are $\lambda=1\times10^5$ and $p=0.03$.
    The upper panel is histogram of the flux loss and the lower panel shows the histogram of noise deterioration.
    The $\mu$ and $\sigma$ values labeled in the figures are the means and standard deviations of their distributions.
}\label{fig:hist_asls_a}
\end{figure}

\begin{figure}
\centering
\includegraphics[width=0.4\textwidth, angle=0]{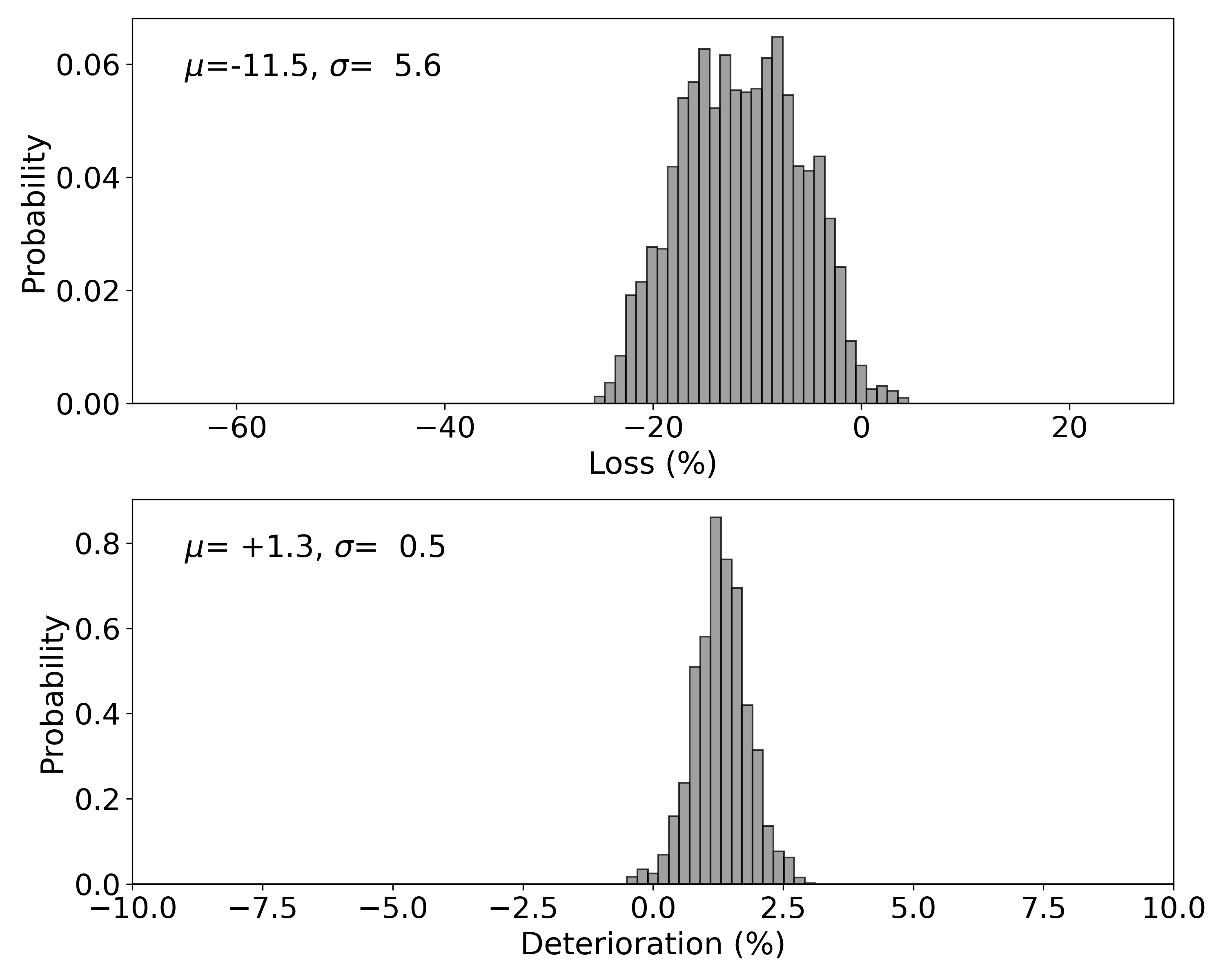}
\caption{
    The distribution of simulation results for Case B using AsLS method with the same optimized parameters for Case A ($\lambda=1\times10^5, p=0.03$).
    The figure instruction follows that is given in Figure\,\ref{fig:hist_asls_a}.
}\label{fig:hist_asls_b}
\end{figure}

The optimized parameters of AsLS is found to be $p=0.03$ and $\lambda=5\times10^5$ for both cases.
We plot histograms of the loss and deterioration factors with the optimized parameters (see Figure\,\ref{fig:hist_asls_a} for Case A and Figure\,\ref{fig:hist_asls_b} for Case B).
The mean flux loss is $-9.6\%$ with a standard deviation of $3.8\%$ for Case A and $-11.5\%$ with $\sigma = 5.6\%$ for Case B.
The rms deterioration distribution has the mean of $1.5\%$ with $\sigma = 0.6$ for Case A and $1.3\%$ with $\sigma = 0.5\%$ for Case B.
This experiment suggests that AsLS can fit the FAST baseline ripples effectively.
However, due to the high noise feature of our data, this method may cause an average line peak intensity loss of $\sim$10\%.


\subsubsection{arPLS}\label{sect:simu:resu:arpls}

\begin{figure}
\centering
\includegraphics[width=0.4\textwidth, angle=0]{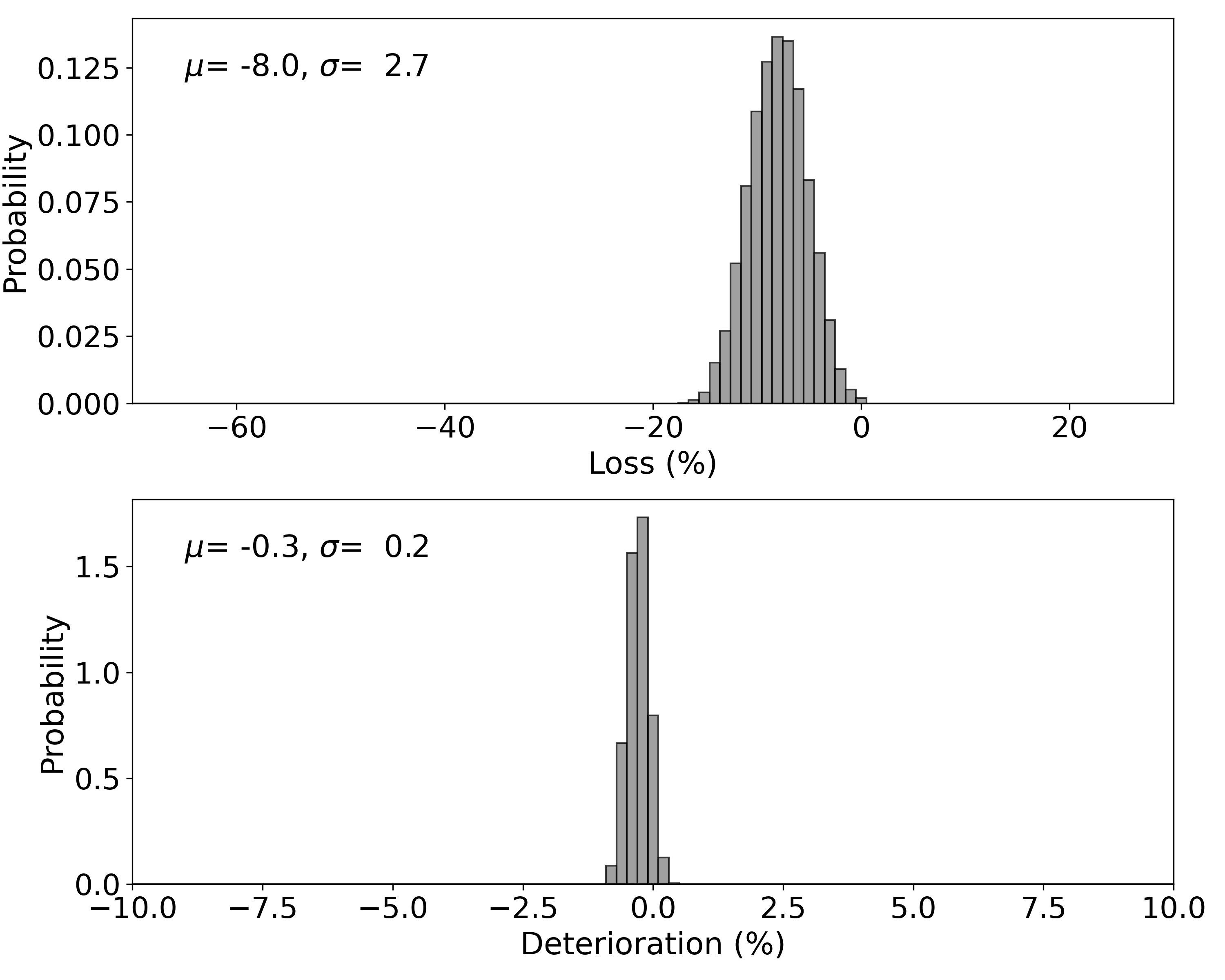}
\caption{
    The distribution of simulation results for Case A using arPLS method with optimized parameter.
    The optimized value of parameter $\lambda$ is $1\times10^6$.
    The figure instruction follows that is given in Figure\,\ref{fig:hist_asls_a}.
}\label{fig:hist_arpls_a}
\end{figure}

\begin{figure}
\centering
\includegraphics[width=0.4\textwidth, angle=0]{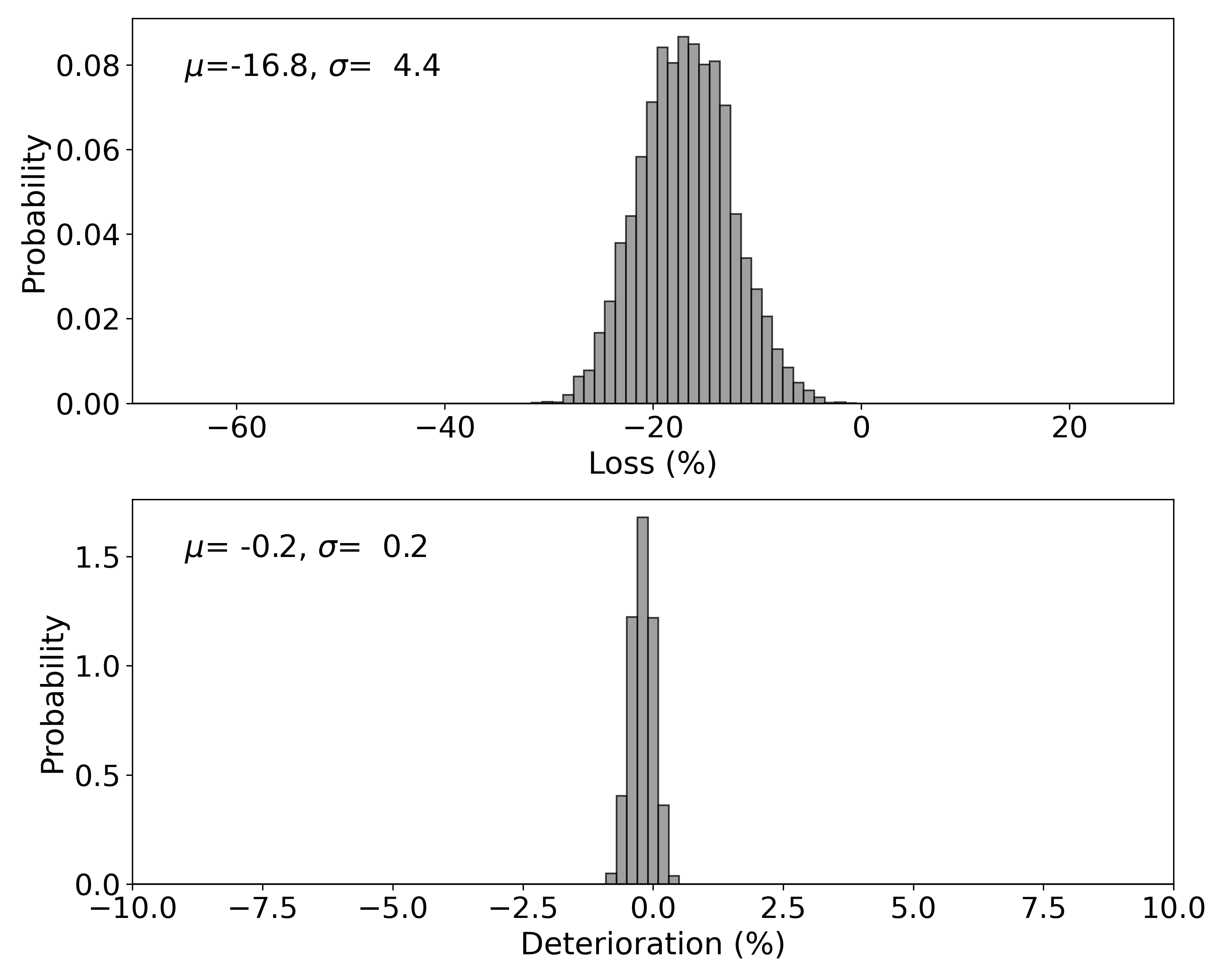}
\caption{
    The distribution of simulation results for Case B using arPLS method with the same optimized parameter for Case A ($\lambda = 1\times10^6$).
    The figure instruction follows that is given in Figure\,\ref{fig:hist_asls_a}.
}\label{fig:hist_arpls_b}
\end{figure}

Adopting the default weight function given by Equation\,\ref{equ:w2}, we obtained the optimized $\lambda=5\times10^6$ for both cases (see Table\,\ref{tab:results}).
To examine the distribution of the results, we also plot the histograms in Figure\,\ref{fig:hist_arpls_a} for Case A and Figure\,\ref{fig:hist_arpls_b} for Case B.

The distribution of flux loss has a mean of $-8.0\%$ with $\sigma$ of $2.7\%$ for Case A and $-16.8\%$ with $4.4\%$ for Case B.
The mean of rms deterioration is $-0.3\%$ with $\sigma = 0.2\%$ for Case A and $-0.2\%$ with $\sigma = 0.2\%$ for Case B.
In comparison with the AsLS results, the smaller value of standard deviation of the flux loss distribution implies that the arPLS method is more stable than AsLS for different conditions.
Although it works better to strong signals as in Case A, it causes more flux loss on average than AsLS for weak signals in Case B.
The negative amplitude of noise deterioration means that the baseline is slightly overfitted.


\subsubsection{asPLS}\label{sect:simu:resu:aspls}

\begin{figure}
\centering
\includegraphics[width=0.4\textwidth, angle=0]{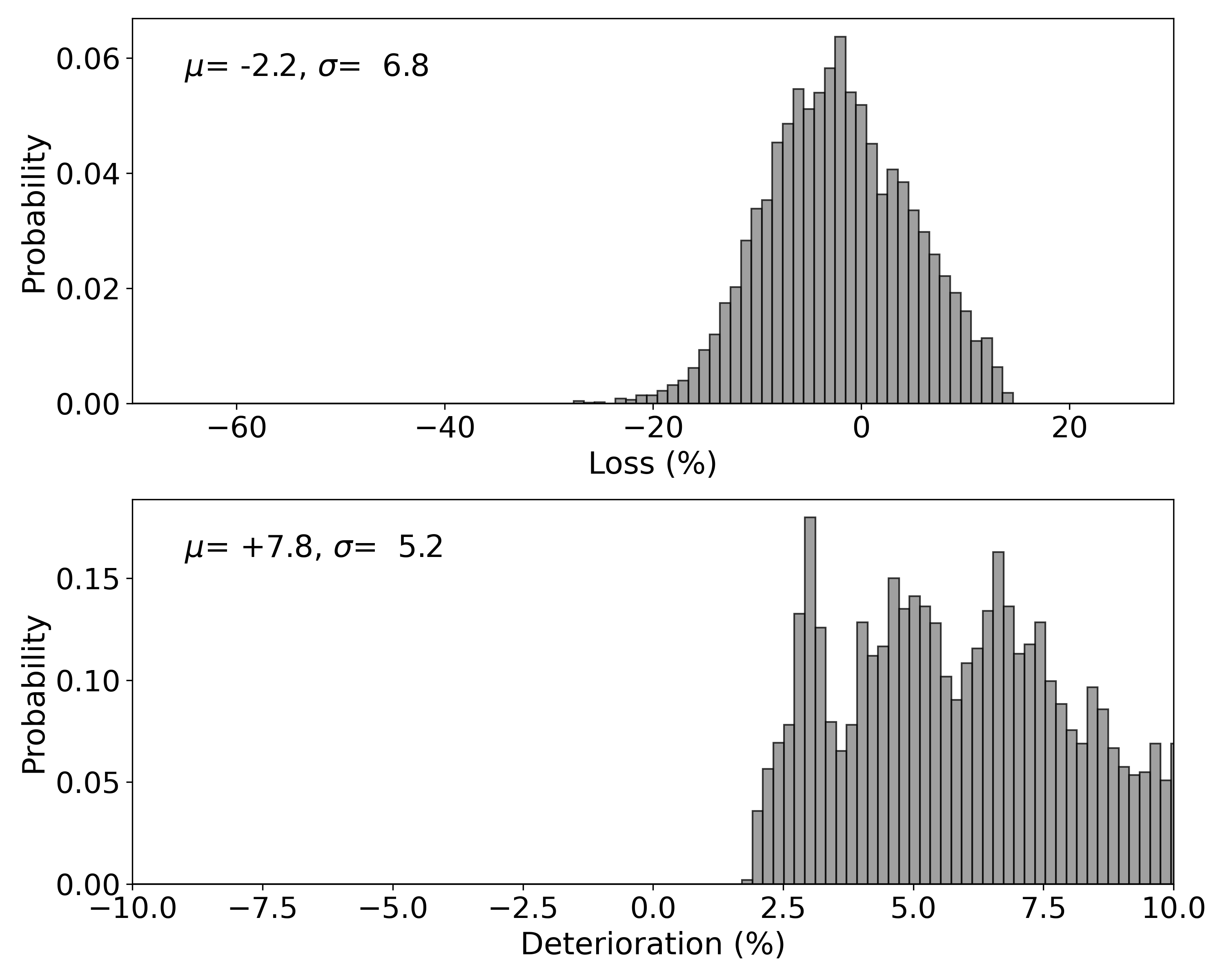}
\caption{
    The distribution of simulation results for Case A using asPLS method with optimized parameters.
    The optimized value of parameter $\lambda$ is $5\times10^5$.
    The figure instruction follows that is given in Figure\,\ref{fig:hist_asls_a}.
}\label{fig:hist_aspls_a}
\end{figure}

\begin{figure}
\centering
\includegraphics[width=0.4\textwidth, angle=0]{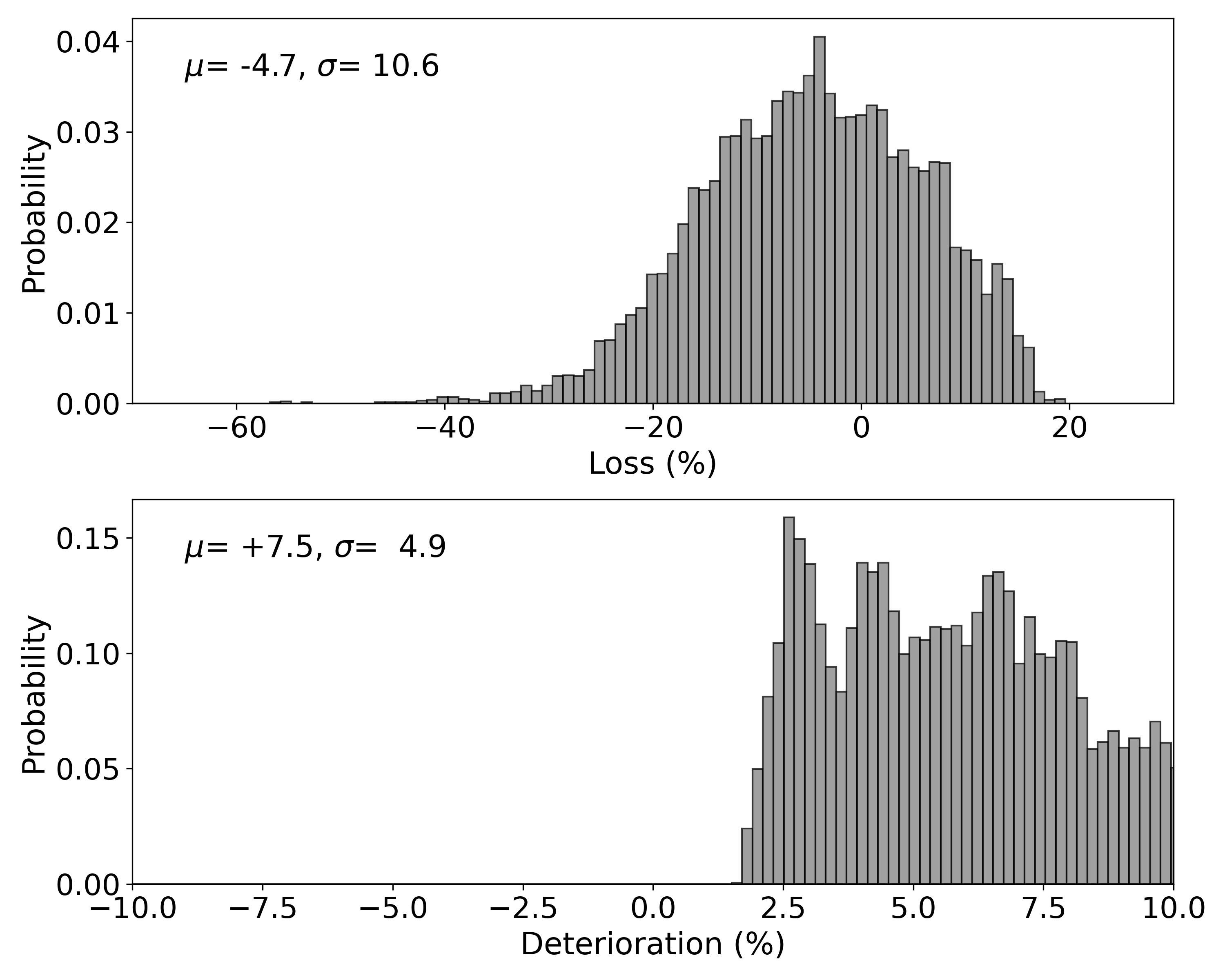}
\caption{
    The distribution of simulation results for Case B using asPLS method with the same optimized parameters for Case A ($\lambda = 5\times10^5$).
    The figure instruction follows that is given in Figure\,\ref{fig:hist_asls_a}.
}\label{fig:hist_aspls_b}
\end{figure}

The smoothing parameter of asPLS is optimized to be $\lambda=5\times10^8$ for both Case A and B (see Table\,\ref{tab:results} for details).
The resulted distributions are also plotted in Figure\,\ref{fig:hist_aspls_a} for Case A and Figure\,\ref{fig:hist_aspls_b} Case B.

The mean of the flux loss distribution is $-2.2\%$ with $\sigma$ of $6.8\%$ for Case A and $-4.7\%$ with $10.6\%$ for Case B.
The mean of rms deterioration is $-3.3\%$ with $\sigma = 3.8\%$ for Case A and $7.5\%$ with $\sigma = 4.9\%$ for Case B.
Comparing with AsLS and arPLS, the flux loss distributes closer to zero although its standard deviation becomes larger.
It seems that the asPLS method could improve the line intensity attenuation problem as expected, but its baseline fitting results may not be very stable for different situations.
Moreover, the spectral rms deteriorates significantly, thus asPLS is not an ideal method for our RRL data reduction.


\subsubsection{rrlPLS}\label{sect:simu:resu:rrlpls}

\begin{figure}
\centering
\includegraphics[width=0.4\textwidth, angle=0]{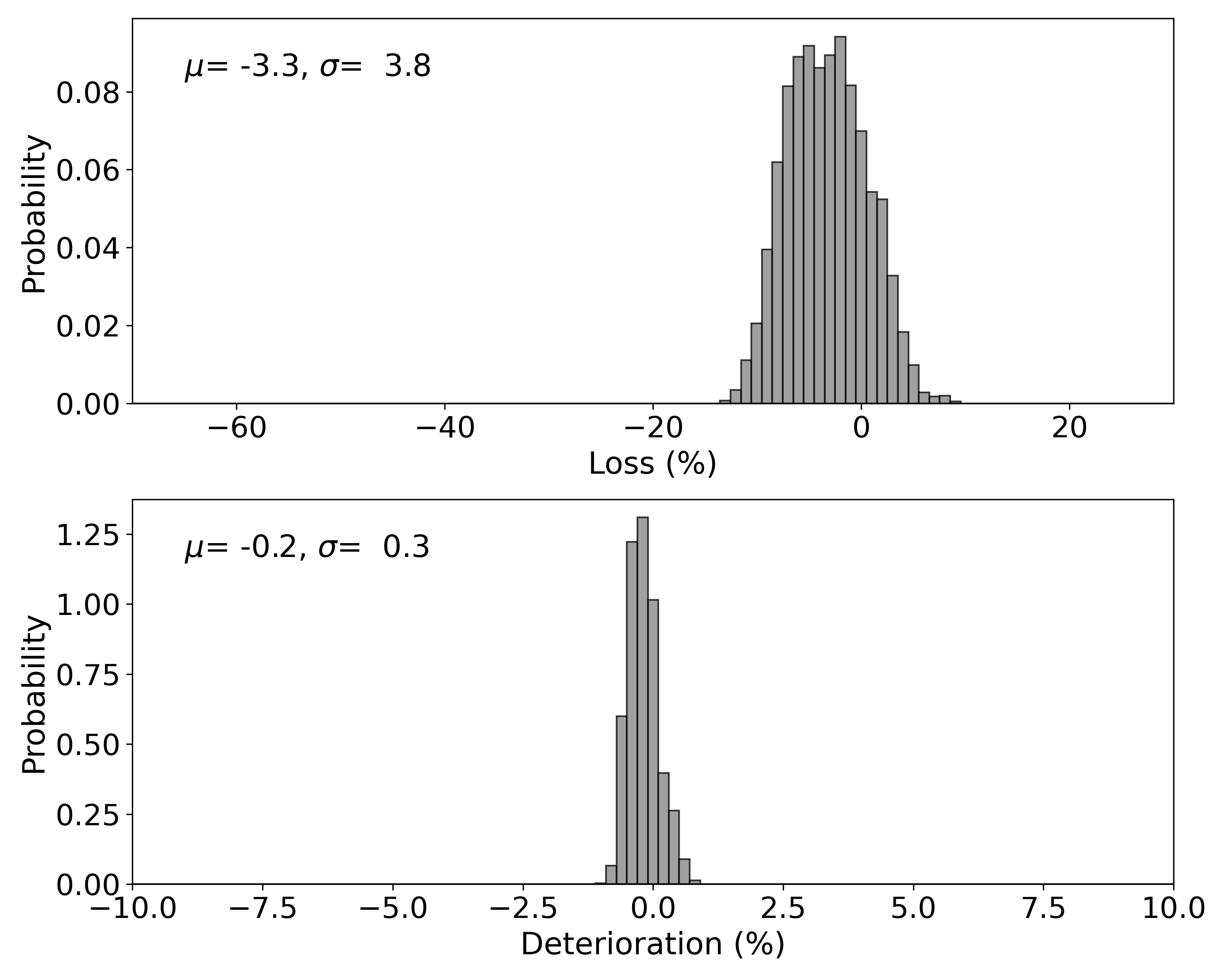}
\caption{
    The distribution of simulation results for Case A using rrlPLS method with optimized parameters.
    The optimized values of parameters are $\lambda = 1\times10^7, k=5,$ and $s=1$.
    The figure instruction follows that is given in Figure\,\ref{fig:hist_asls_a}.
}\label{fig:hist_rrlpls_a}
\end{figure}

\begin{figure}
\centering
\includegraphics[width=0.4\textwidth, angle=0]{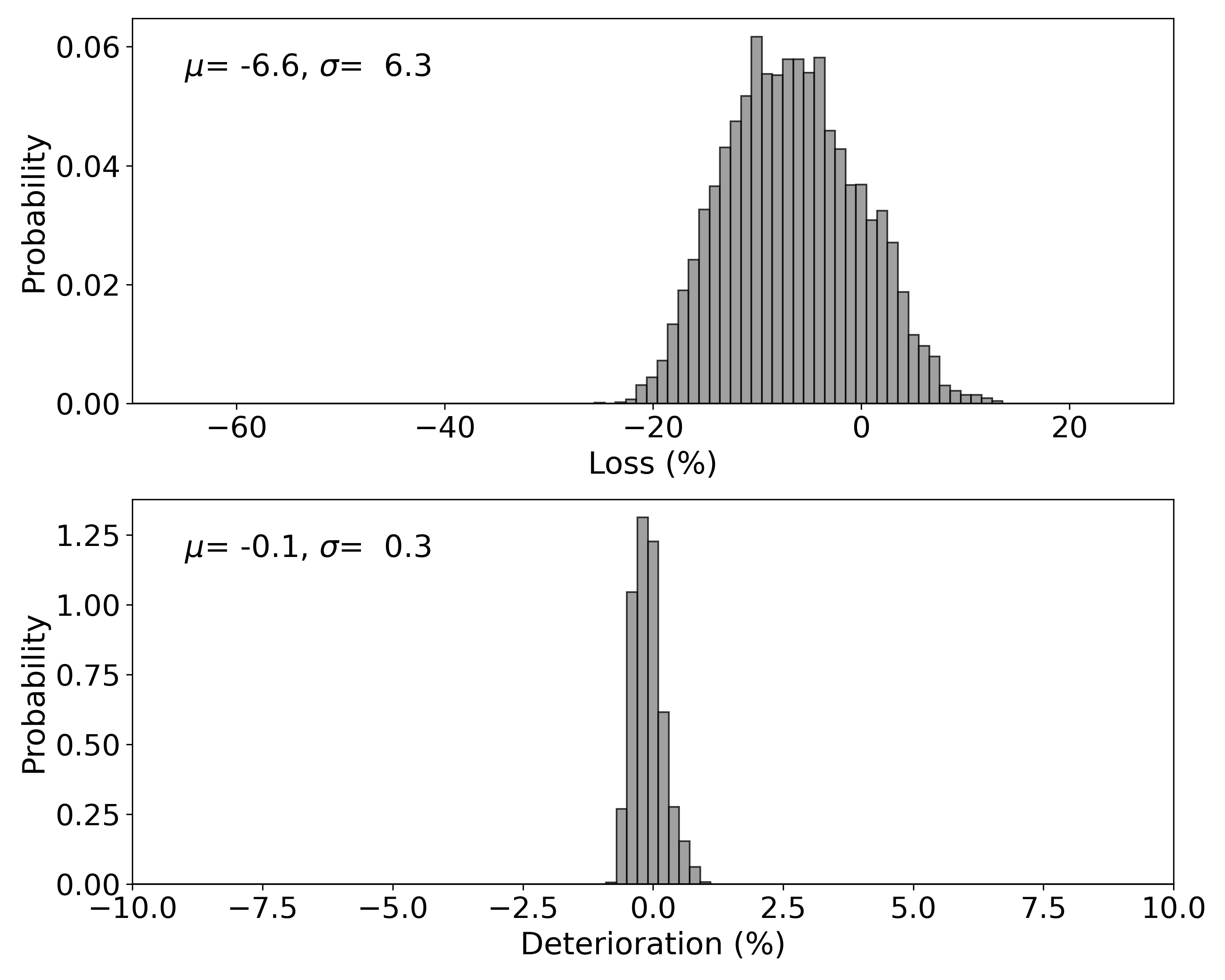}
\caption{ The distribution of simulation results for Case B using rrlPLS method with the same optimized parameters for Case A ($\lambda = 1\times10^7, k=5,$ and $s=1$).
    The figure instruction follows that is given in Figure\,\ref{fig:hist_asls_a}.
}\label{fig:hist_rrlpls_b}
\end{figure}

Similar as other methods, simulations with rrlPLS are conducted.
When $\lambda=1\times 10^7$, we obtain the best baseline fitting results.
Figure\,\ref{fig:hist_rrlpls_a} and \,\ref{fig:hist_rrlpls_b} present the histograms of the results for Case A and B.

The mean of the flux loss distribution is $-3.3\%$ with $\sigma$ of $3.8\%$ for Case A and $-6.6\%$ with $6.3\%$ for Case B.
The mean of rms deterioration is $-0.2\%$ with $\sigma = 0.3\%$ for Case A and $-0.1\%$ with $\sigma = 0.3\%$ for Case B.
Comparing to the other three methods, the $\sim5\%$ flux loss introduced with nearly $\sim0\%$ noise deteriorations make rrlPLS the most promising baseline correction method to our RRL spectra.



\section{Apply to real RRL data}\label{sect:rrl}
The four methods discussed in Section\,\ref{sect:pls} and \ref{sect:simu} are employed to fit the baselines of observed raw spectra.
We extract a spectral segment of H$169\alpha$ from the full bandpass of calibrated data.
Then the spectra of the RRL segment are baseline removed using one of these methods.
Finally the processed spectra are re-grided into data cubes.

For each method, the optimized parameters listed in Table\,\ref{tab:results} are firstly adopted.
Considering the difference between the simulated and observed data sets, we further tuned the parameters by small steps.
No clear improvements have been seen except for the rrlPLS, in which $\lambda=2\times10^8$ is configured instead of $1\times10^7$.

\begin{figure*}
\centering
\begin{subfigure}{0.5\textwidth}
    \centering
    \includegraphics[width=0.9\linewidth, angle=0]{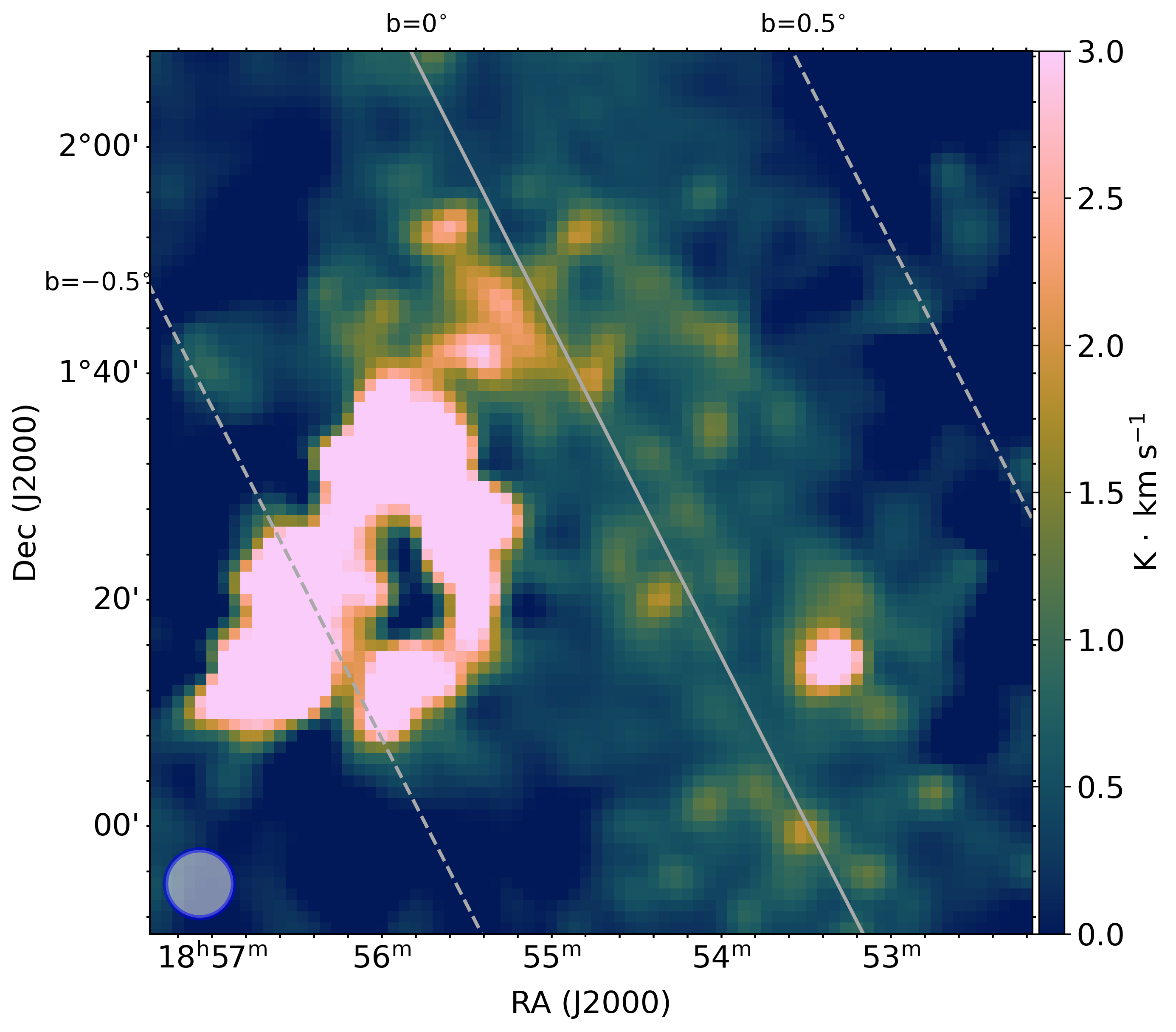}
    \caption{SIGGMA}
    \label{fig:siggma}
\end{subfigure}
\begin{subfigure}{.5\textwidth}
    \centering
    \includegraphics[width=0.9\linewidth, angle=0]{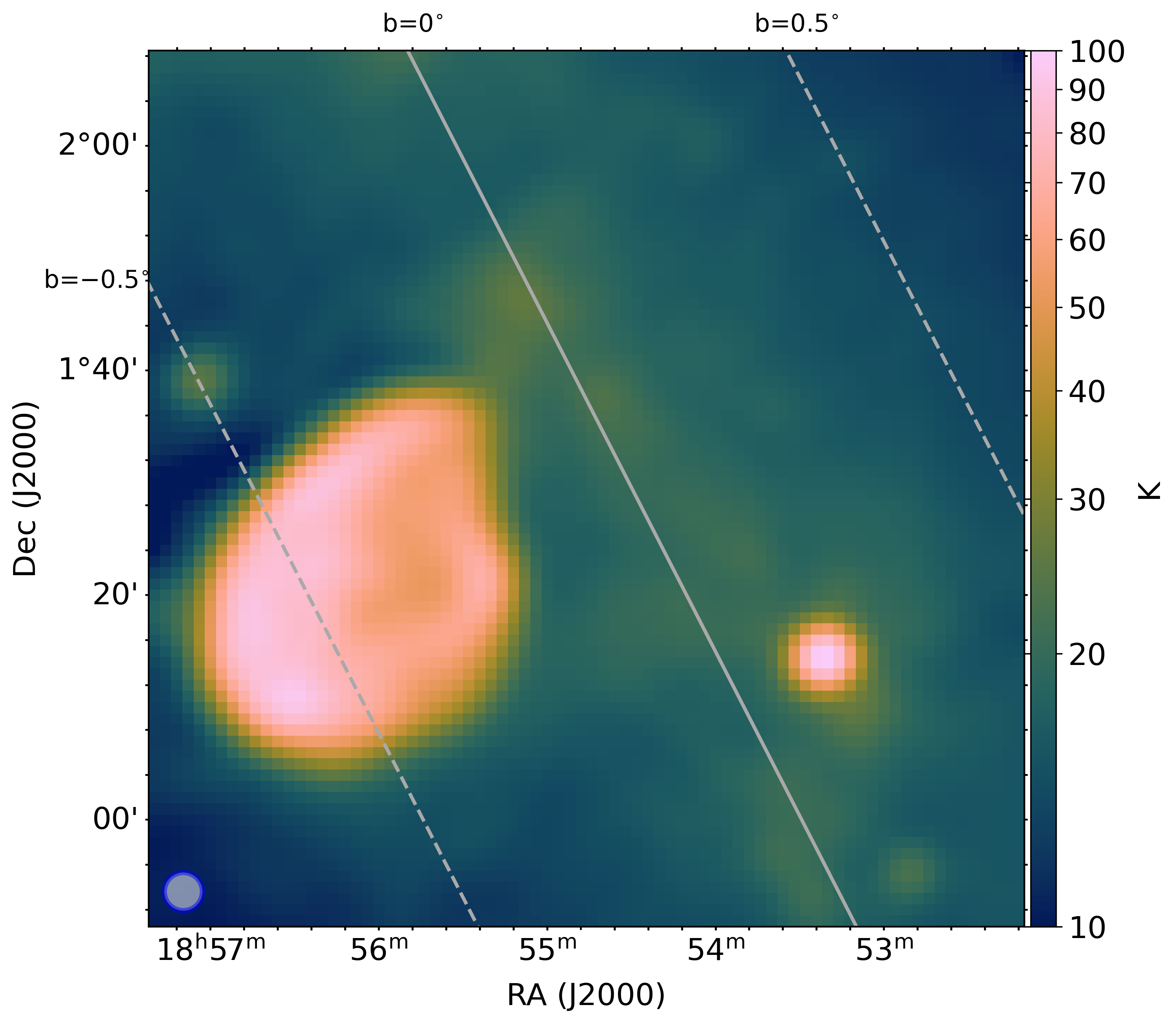}
    \caption{VGPS}
    \label{fig:vgps}
\end{subfigure}
\caption{
    (a) The SIGGMA RRL 0$^{\mathrm{th}}$ moment map integrated over the velocity range from 20 to 100\,\kms \citep{Liu2019}. 
    The blue circle at the bottom left corner shows the SIGGMA resolution of 6\arcmin. 
    (b) The VGPS continuum map at 1.4\ghz\citep{Stil2006}.
    The VGPS data is convolved to FAST HPBW of 3\arcmin\;at 1350\mhz (blue circle at the lower left corner).
    Both images are re-projected to match with the FAST image grid. The bright extended source located at the middle east in the field is the supernova remnant W44, who shows strong non-thermal continuum emission.
}\label{fig:compare}
\end{figure*}

\begin{figure*}
\centering
\includegraphics[width=0.7\textwidth, angle=0]{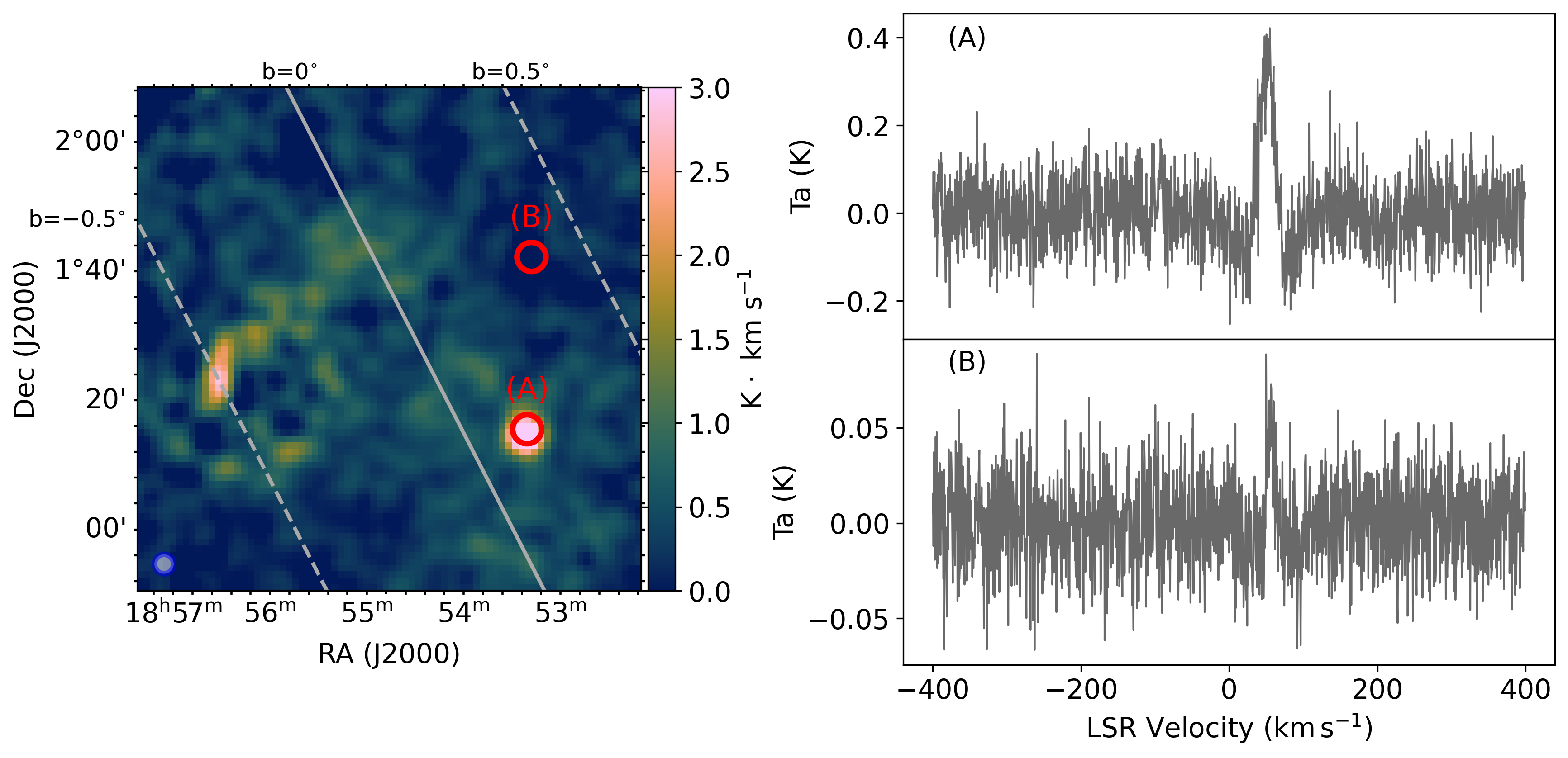}
\caption{
    The results of H169$\alpha$ processed using AsLS method.
    The image on the left is the 0$^{\mathrm{th}}$ moment map integrated over the velocity range from 20 to 100\,\kms from the cube.
    The red circles marked as (A) and (B) in the map are locations with strong and weak RRLs.
    The blue circle at the bottom left corner shows the FAST beam size of 3\arcmin. 
    The right panel plot two spectra at the locations marked as (A) and (B) in the left-hand moment map.
    (A) is apart from strong continuum source, where RRL signal is weak. 
    (B) is a known bright H\,$_{\mathrm{II}}$ region, who shows intensive RRL emission.
    The spectra of (A) and (B) are corresponding to the Case A and B in the simulation.
}
\label{fig:asls}
\end{figure*}

\begin{figure*}
\centering
\includegraphics[width=0.7\textwidth, angle=0]{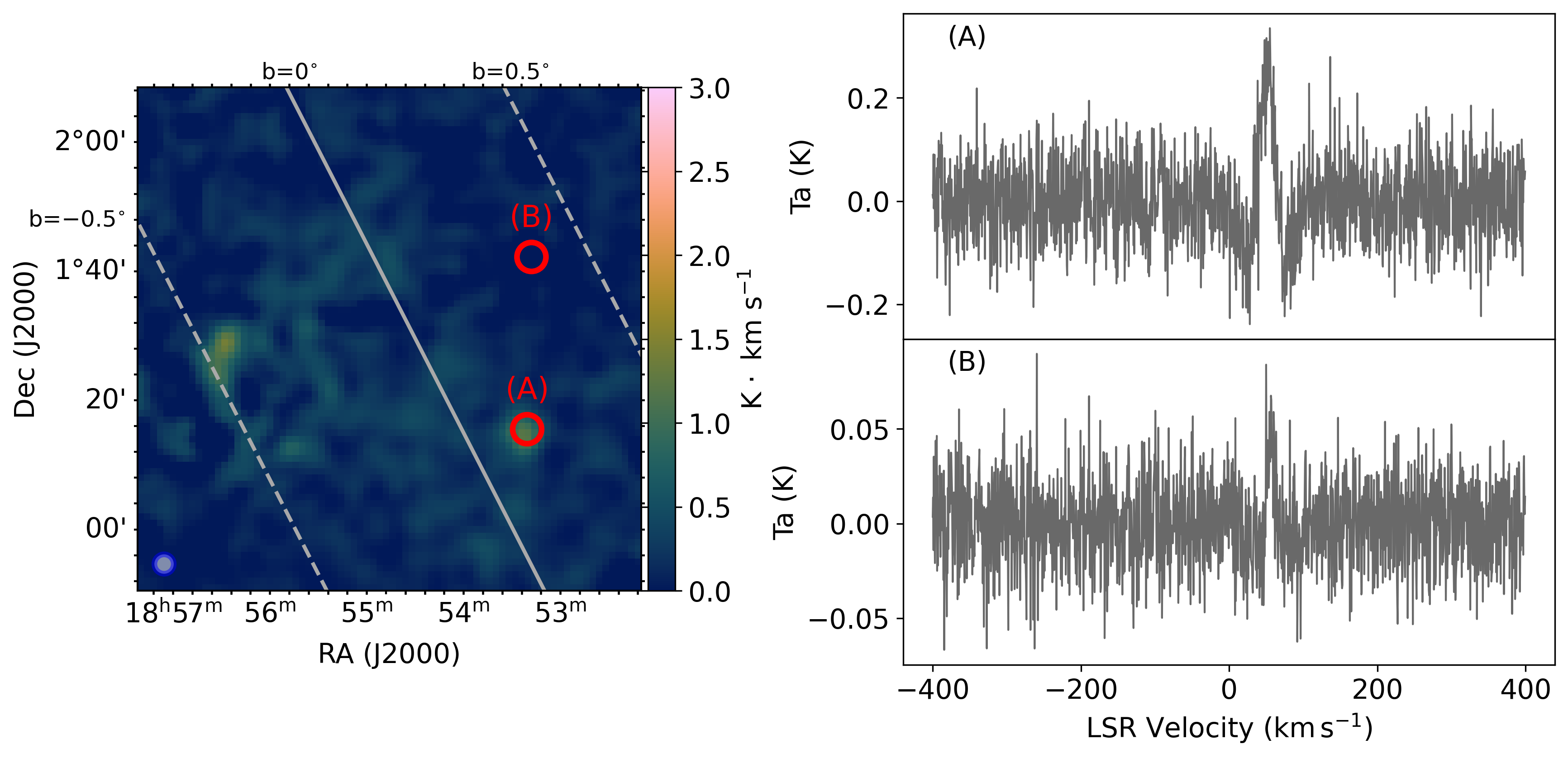}
\caption{
    The results of H169$\alpha$ processed using arPLS method.
    The figure instruction follows that is given in Figure\,\ref{fig:asls}.
}\label{fig:arpls}
\end{figure*}

\begin{figure*}
\centering
\includegraphics[width=0.7\textwidth, angle=0]{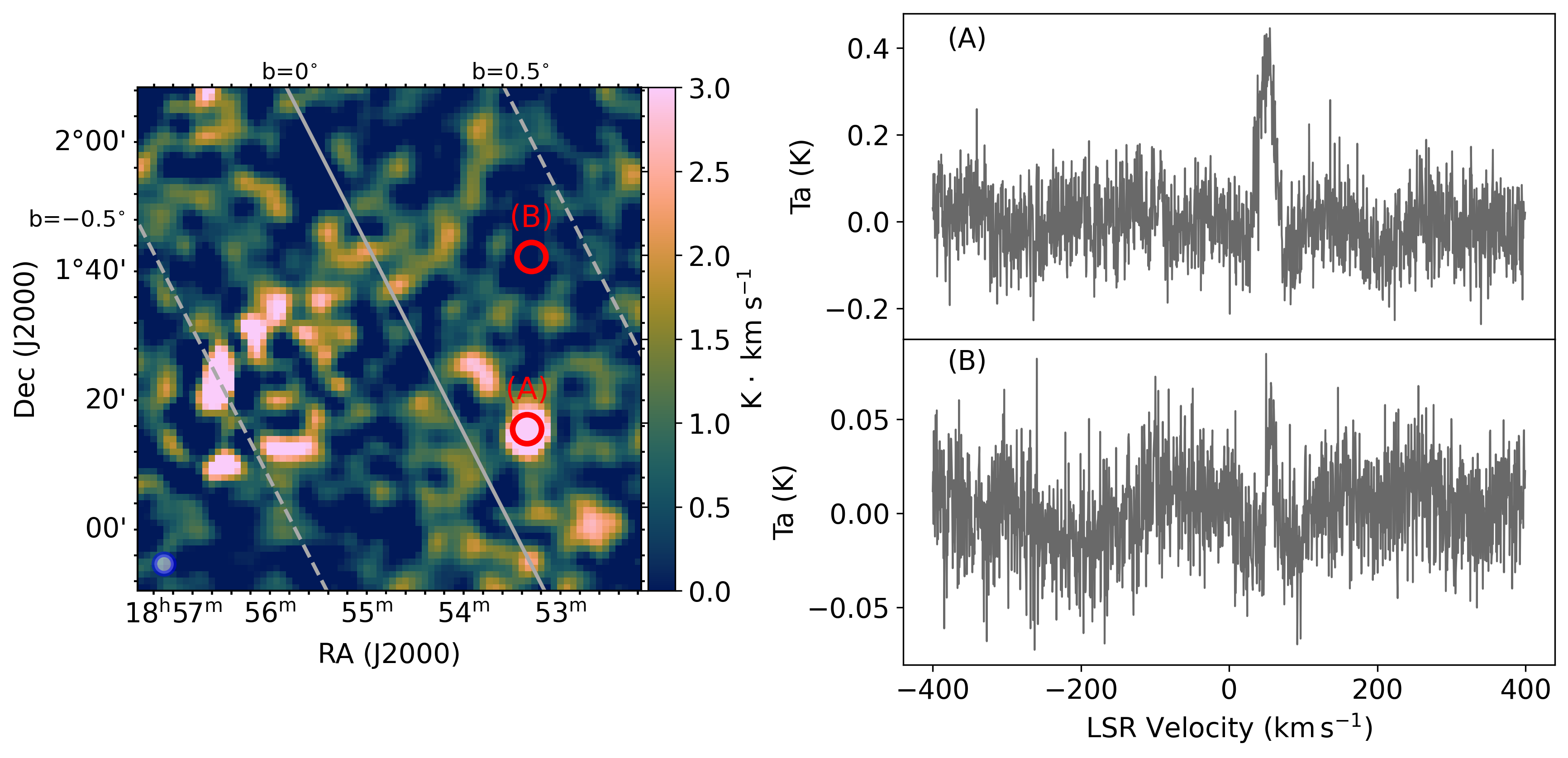}
\caption{
    The results of H169$\alpha$ processed using asPLS method.
    The figure instruction follows that is given in Figure\,\ref{fig:asls}.
}\label{fig:aspls}
\end{figure*}

\begin{figure*}
\centering
\includegraphics[width=0.7\textwidth, angle=0]{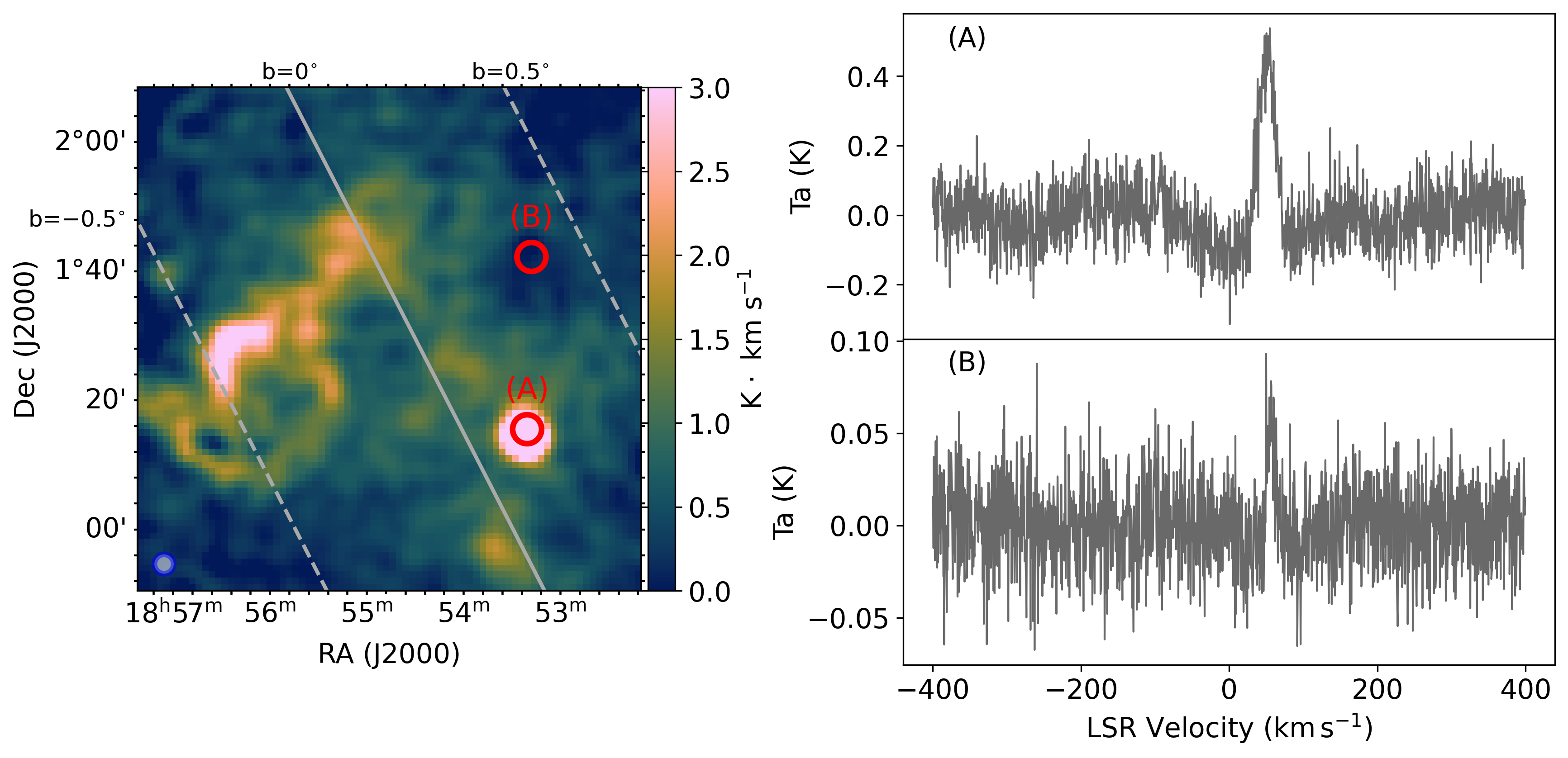}
\caption{
    The results of H169$\alpha$ processed using rrlPLS method.
    The figure instruction follows that is given in Figure\,\ref{fig:asls}.
}\label{fig:rrlpls}
\end{figure*}

To compare the FAST RRL results with previous studies, we provide the RRL 0$^{\mathrm{th}}$ moment map (Figure\,\ref{fig:siggma}) given by the Survey of Ionized Gas in the Galaxy Made with Arecibo \citep[SIGGMA,][]{Liu2013}.
The sensitivity of stacked RRLs from SIGGMA is remarkable, however, there may be some unreliable spatial-extended features in the map due to its survey strategy and data quality \citep{Liu2019}.
Therefore, we also present the 1.4\ghz continuum map (Figure\,\ref{fig:vgps}) from the VLA Galactic Plane Survey \citep[VGPS,][]{Stil2006}.
For a better comparison, the VGPS map was convolved to the FAST HPBW and re-projected to the grid of FAST RRL maps.
The data processing results using AsLS, arPLS, asPLS, and rrlPLS methods are presented in Figure\,\ref{fig:asls}, \ref{fig:arpls}, \ref{fig:aspls}, and \ref{fig:rrlpls}.
In each figure, the left panel presents the 0$^{\mathrm{th}}$ moment map integrated over the velocity range between 20 and 100\,\kms from the cube.
The right panel gives two spectra at the locations marked as (A) and (B) in the moment map.
(A) is a known bright \hii region, where intensive RRL emission exists.
(B) is a relatively `empty' spot in the field, where no strong RRL signals are expected. 
The spectra of (A) and (B) are corresponding to the Case A and B in the simulation.

The AsLS method was first adopted for our project.
In Figure\,\ref{fig:asls}, one can see smoothed gas structures and clean spectral baselines.
However, it introduces notable flux loss to strong emitting sources.
The arPLS method gives the best baseline fitting but also causes the most severe flux losses.
Most of the RRL emissions are eliminated as demonstrated in Figure\,\ref{fig:arpls}.
The asPLS method was designed to retain line signals from noisy spectra.
Although the corresponding flux loss is small, the resulting baseline quality is the worst comparing to the other three methods (Figure\,\ref{fig:aspls}).
Furthermore, because of the bad baseline, the line peak intensities of weak sources are inaccurate.

Finally, the rrlPLS method presents the most promising results in Figure\,\ref{fig:rrlpls}.
Comparing the spectrum of (A) in Figure\,\ref{fig:rrlpls} with that in Figure\,\ref{fig:aspls}, the line peak intensities are identical. 
The flux loss introduced by rrlPLS is as small as asPLS.
Whereas the baseline of strong continuum source given by rrlPLS is not as good as that from AsLS or arPLS, and those of weak emissions are similar.
Furthermore, the 0$^{\mathrm{th}}$ moment map resulted from rrlPLS reveals the most intensive gas structures than that from other methods, implying that rrlPLS with $\lambda=2\times10^8$ produces the best RRL data sets.

The results of RRL maps processed by the PLS methods agree with the simulation results in Section\,\ref{sect:simu}.
Therefore, discussions on emission line searching, gas kinematics from LSR velocities, and gas morphology from relative line peak intensities are reliable.
One should still be cautious about the uncertainties for further astrophysical analysis using the line profiles.
Before line profile fitting and further calculations for individual strong continuum sources, a high order polynomial baseline removal is suggested by masking the velocity range of the detected RRL.


\section{Evaluation with fake source injection}\label{sect:fake}
Finally, it is necessary to deploy a quantitative analysis on the line distortions through the whole process using rrlPLS.
Since the true intensities of observed RRLs is unknown, simulated Gaussian profiles are added to the raw spectra centering at LSR velocity of $-$300\,\kms so that fake signals will not overlap with real RRLs.
To imitate a point source, all spectra were injected with weights according to FAST beam pattern and the angular distances from positions where spectra were obtained to the fake source location.
To cover different baseline scenarios, three sources, located at strong, medium, and weak continuum background, were simulated and added to raw data sets.
Data cube was then produced using our pipeline, from which the spectra of the fake sources were extracted and fitted.

\begin{figure*}
\centering
\includegraphics[width=0.75\textwidth, angle=0]{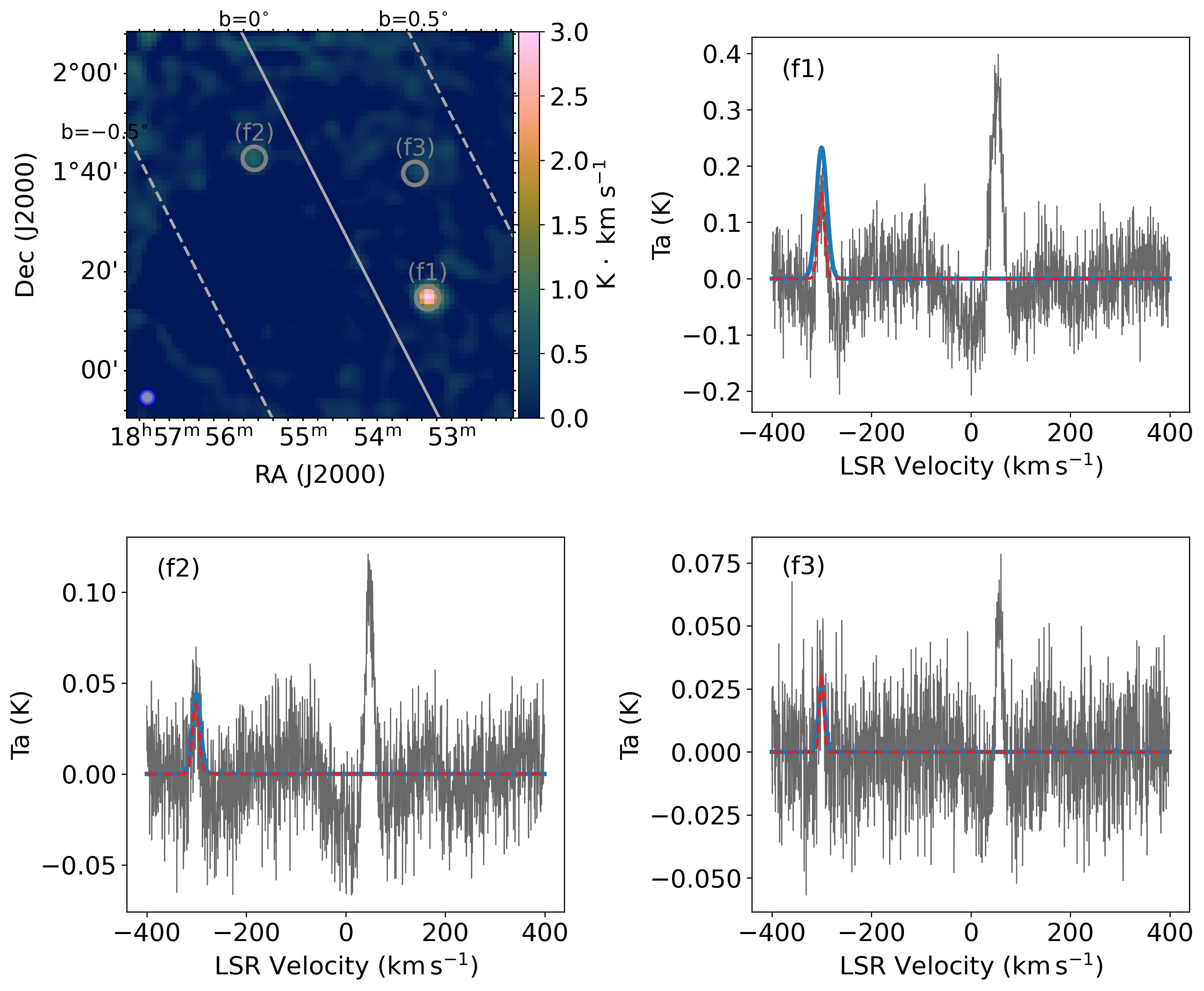}
\caption{
    The comparison of rrlPLS fitting results with simulated spectra injected into H169$\alpha$ data.
    The top left is the 0$^{\mathrm{th}}$ moment map integrated over the velocity range from $-$320 to $-$280\,\kms, within which the fake line profiles are injected.
    The blue circle at the bottom left corner of the map shows the FAST beam size of 3\arcmin.
    The top right and bottom plots are the spectra extracted from data cube towards the fake sources.
    The solid gray lines are the processed spectra, solid blue lines are the injected Gaussian profiles, and dashed red lines are the fitted line profiles to the spectra.
}
\label{fig:fakerrl}
\end{figure*}

Figure\,\ref{fig:fakerrl} shows the 0$^{\mathrm{th}}$ moment map and the spectra of the three fake sources, labeled as (f1), (f2), and (f3).
(f1) is located at a known bright \hii region (strong continuum background), (f2) is within a extended gas structure (medium continuum background), and (f3) is at a weak emission spot (weak continuum background).
The simulated (solid blue) and fitted (dashed red) line profiles are overlaid.
For relatively weak (f2) and (f3), the simulated and processed line intensities are identical.
But for the stronger (f1), the processed profile is notably weaker than the simulated.

To compare with the traditional method, baseline fitting with a 3$^{\mathrm{rd}}$ order polynomial (Poly-3), with a velocity mask from $-320$ to $-280$\kms, were also carried out besides rrlPLS.
Table\,\ref{tab:fake} presents the comparison of simulated and processed line profiles using both methods.
Although they follows same trends, the results of rrlPLS are more consistent than that of Poly-3.
The latter does not reduce the standing wave ripples in the spectra, thus will not generate reliable emission structures in the map.
The flux losses of (f2) and (f3) of rrlPLS are possibly introduced by the rms noise of spectra.
After a 5$^{\mathrm{th}}$ order polynomial baseline fitting to the (f1) spectrum from rrlPLS, the fitted line intensity is well recovered.

To summarize, line profiles with low signal-to-noise ratio were less affected by rrlPLS, whose uncertainty were mainly caused by the rms noise.
Although the majority of the detected RRLs are weak, to which the rrlPLS method is acceptable, still the strong line peaks may affect the baseline fitting towards a few intensive positions in the field.
Therefore, after the identification of strong RRLs, high order polynomial baseline removals are suggested for accurate line profile fittings.
In addition, we note that the line widths are reduced after baseline processing from both Poly-3 and rrlPLS.
Since the line narrowing is not distinct from the methods applied, it may be caused by the remained baseline ripples.
For scientific discussion with the current data sets, one need to be careful about calculations using line widths, which may be under estimated.

\begin{table*}
    \caption{The comarison of rrlPLS fitting resuts with simulated spectra injected into real RRL data.}
\label{tab:fake}
\centering
\begin{threeparttable}
    \begin{tabular}{lccccccccc}
        \hline\hline
        \multirow{2}{*}{Source} & \multicolumn{3}{c}{True Profile}& \multirow{2}{*}{Method} & \multicolumn{3}{c}{Fitted Profile} & \multirow{2}{*}{Flux Loss} & \multirow{2}{*}{rms Noise}\\
        \cline{2-4}\cline{6-8}
                            & Peak                   & V$_{\mathrm{LSR}}$      & FWHM                &        & Peak            & V$_{\mathrm{LSR}}$ & FWHM         &         &       \\
                            &  (K)                   & \kms                    & \kms                &        &  (K)            & \kms               & \kms         & ($\%$)  & (K)   \\
        \hline
        \multirow{2}{*}{f1} & \multirow{2}{*}{0.233} & \multirow{2}{*}{$-$300} & \multirow{2}{*}{25} & rrlPLS & 0.152$\pm$0.019 & $-$300.8$\pm$1.0   & 15.6$\pm$1.0 & $-$34.8 & 0.050 \\ 
                            &                        &                         &                     & Poly-3 & 0.139$\pm$0.032 & $-$301.9$\pm$1.6   & 14.5$\pm$1.6 & $-$40.3 & 0.061 \\
        \multirow{2}{*}{f2} & \multirow{2}{*}{0.044} & \multirow{2}{*}{$-$300} & \multirow{2}{*}{20} & rrlPLS & 0.039$\pm$0.007 & $-$301.6$\pm$1.3   & 15.9$\pm$1.3 & $-$11.4 & 0.019 \\
                            &                        &                         &                     & Poly-3 & 0.040$\pm$0.008 & $-$302.9$\pm$1.7   & 16.9$\pm$1.7 & $-$9.1  & 0.021 \\
        \multirow{2}{*}{f3} & \multirow{2}{*}{0.026} & \multirow{2}{*}{$-$300} & \multirow{2}{*}{15} & rrlPLS & 0.031$\pm$0.006 & $-$300.1$\pm$1.0   & 10.5$\pm$1.0 & $+$19.2 & 0.018 \\ 
                            &                        &                         &                     & Poly-3 & 0.033$\pm$0.006 & $-$299.9$\pm$1.1   & 10.8$\pm$1.1 & $+$26.9 & 0.018 \\
        f1$^{\mathrm{a}}$   & 0.233 & $-$300 & 25 &rrlPLS &0.216$\pm$0.010 & $-$300.4$\pm$0.5 & 21.6$\pm$0.5 & $-$7.1  & 0.050 \\ 
        \hline
    \end{tabular}
    \begin{tablenotes}
    \item Col.1 are the name of injected fake sources.
        Cols.2-4 list the true profile parameters of the simulated spectra.
        Col.5 gives the baseline removal methods applied in the pipeline.
        `Poly-3' stands for the 3$^{\mathrm{rd}}$ order polynominal fitting.
        Col.6-8 list the fitting parameters of the simulated spectra after the baseline correction processed.
        Col.9 and 10 are the relative flux loss and the rms values of spectra, which are calculated from velocity range between $+$300 and $+$400\kms.
    \item$^{\mathrm{a}}$ The line profile is fitted after an extra 5th order polynomial with velocity mask covering $-$320 to $-$280\kms.
    \end{tablenotes}
\end{threeparttable}
\end{table*}



\section{Conclusions} \label{sect:con}

To investigate the ionized environment in the Galaxy using FAST, RRL map of 1 deg$^2$ on the Galactic plane has been processed, which serves as a pilot study for a further large-scale Galactic plane RRL survey with FAST.
In this paper, We introduced the observing details, survey configurations, and data processing pipeline developments.
The data shows that the frequency bandpass given by the FAST 19 beam L-band receiver is severely affected by RFIs and standing wave ripples, which brings a major challenge lying in the baseline fitting step.
Low order polynomial baseline removal method, which is widely used for spectroscopy studies in radio astronomy, is not suitable for this complex scenarios.

To solve the baseline problem, we investigate a series of PLS-based baseline correction methods in this paper.
The AsLS, arPLS, and asPLS methods were evaluated using simulated spectra according to the actual features of FAST bandpass.
To further improve the results of baseline correction, we developed a modified method, rrlPLS, by adopting the advantages of arPLS and asPLS.
Optimized parameters were obtained from our simulations.
The four methods were then applied and compared by reducing the real FAST data.
The rrlPLS with the optimized parameter $\lambda=2\times10^8$ revealed the most sensitive and reliable RRL emission features in the 0$^{\mathrm{th}}$ moment map, and thus was well-suited for our project.

We finally verified the line distortion, which the rrlPLS method may cause, using fake RRL sources injected into the raw data sets.
Small distortions were identified by comparing the processed profiled to the simulated ones. 
It is concluded that the weaker the line intensity the less it may be affected by this baseline removal method.
While for strong emission regions, an extra high order polynomial is suggested before fitting the line profile accurately, so that the uncertainty is mainly introduced by the rms noise. 
Note that the fitted line width may be smaller than the actual signal due to the baseline ripples and the current limitation of the method.
Cautious considerations should be made for line width related science cases.

A Galactic plane RRL mapping project with FAST is now in progress.
The baseline correction technique verified in this paper will be applied in our future data processing pipeline.

\section*{Acknowledgements}

This work is supported by the National Key Basic Research and Development Program of China(grant No. 2017YFA0402604 and 2018YFA0404703), and the National Natural Science Foundation of China (No.11703048) and the Open Project Program of the Key Laboratory of FAST, NAOC, Chinese Academy of Sciences.
 This work made use of the data from FAST (Five-hundred-meter Aperture Spherical radio Telescope). FAST is a Chinese national mega-science facility, operated by the National Astronomical Observatories, Chinese Academy of Sciences.



\bibliographystyle{pasa-mnras}
\bibliography{ref} 


\end{document}